\documentclass[showkeys,showpacs,superscriptaddress]{revtex4}
\usepackage{amsmath}
\usepackage{amssymb}
\usepackage{graphicx}
\usepackage{color}
\usepackage{longtable}
\usepackage{hyperref}

\usepackage{makecell}
\begin{document}

\title{
Energy spectrum and the mass gap\\ from nonperturbative quantization \`a la Heisenberg
}

\author{
Vladimir Dzhunushaliev
}
\email{v.dzhunushaliev@gmail.com}
\affiliation{
Department of Theoretical and Nuclear Physics,  Al-Farabi Kazakh National University, Almaty 050040, Kazakhstan
}
\affiliation{
Institute of Experimental and Theoretical Physics,  Al-Farabi Kazakh National University, Almaty 050040, Kazakhstan
}
\affiliation{
Institute of Physicotechnical Problems and Material Science of the NAS of the Kyrgyz Republic, 265 a, Chui Street, Bishkek 720071, Kyrgyzstan
}
\affiliation{
Institute of Systems Science,
Durban University of Technology, P. O. Box 1334, Durban 4000, South Africa
}

\author{Vladimir Folomeev}
\email{vfolomeev@mail.ru}
\affiliation{
Institute of Experimental and Theoretical Physics,  Al-Farabi Kazakh National University, Almaty 050040, Kazakhstan
}
\affiliation{
Institute of Physicotechnical Problems and Material Science of the NAS of the Kyrgyz Republic, 265 a, Chui Street, Bishkek 720071, Kyrgyzstan
}


\begin{abstract}
Using approximate methods of nonperturbative quantization \`a la Heisenberg and taking into account the interaction of gauge fields with quarks, we find regular solutions describing the following configurations: (i) a spinball consisting of two virtual quarks with opposite spins; (ii) a quantum monopole; (iii) a spinball-plus-quantum-monopole system; and (iv) a spinball-plus-quantum-dyon system.
A comparison with quasi-particles obtained by lattice and phenomenological analytical calculations is carried out.
All these objects (except the spinball) are embedded in a bag created by the quantum coset condensate
consisting of the SU(3)/(SU(2)~$\times$~U(1)) gauge fields. The existence of these objects is due to the Meissner effect,  which implies that the SU(2)~$\times$~U(1) gauge fields are expelled from the condensate.
The physical interpretation of these solutions is proposed in two different forms: (i) an approximate glueball model;  and (ii) quantum fluctuations in the coset condensate of the nonperturbative vacuum or in a quark-gluon plasma.
For the spinball and the spinball-plus-quantum-monopole configuration, we obtain
energy spectra, in which mass gaps are present. The process of deconfinement is discussed qualitatively. It is shown that the quantum chromodynamics constant $\Lambda_{\text{QCD}}$ appears in the nonperturbative quantization \`a la Heisenberg as some constant controlling the correlation length of quantum fields in a spacelike direction.
\end{abstract}



\maketitle

\section{Introduction}

In quantum chromodynamics (QCD), there is a consensus that the structure of the QCD vacuum is much more complicated than that of quantum electrodynamics. It is generally believed that quantum fluctuations having the form of a monopole (a hedgehog) do exist in the QCD vacuum. But the radial magnetic field of a monopole in the QCD vacuum decreases asymptotically according to an exponential law, unlike the magnetic field of the 't~Hooft-Polyakov monopole which decreases according to a power law. This is a crucial distinction, leading to the problem of obtaining a monopole with a magnetic field that decays exponentially at infinity (a hedgehog).

Here, we employ a nonperturbative quantization  method \`{a} la Heisenberg
and its three-equation approximation for obtaining a set of equations describing a quantum monopole/dyon interacting with virtual quarks and having the required asymptotic behavior. We also show that in such system there exist solutions describing (i) virtual quarks (a spinball); (ii) a quantum monopole;
(iii) a spinball-plus-quantum-monopole configuration; and (iv) a spinball-plus-quantum-dyon system, and the configurations (ii)-(iv) are enclosed within a bag. Also, in the systems (ii)-(iv), there takes place the Meissner effect: color magnetic and electric fields are expelled by the condensate created by the SU(3)/(SU(2)~$\times$~U(1)) gauge fields.

The three-equation approximation suggested by us is a gauge noninvariant approximation within which an infinite set of Dyson-Schwinger equations
is cutted off to first three equations. The gauge noninvariance consists in that (i)~we divide
  SU(3)  degrees of freedom of a gauge field into
 SU(2)~$\times$~U(1) and coset  SU(3)/(SU(2)~$\times$~U(1)) degrees of freedom and
 (ii)~there appears a mass term $ \left( \mu^2 \right)^{ab \mu \nu} $ in Eq.~\eqref{1-10} for a massive
  SU(2)~$\times$~U(1) gauge field.

At the present time, there is a universally accepted point of view according to which the vacuum in gluodynamics and chromodynamics is described fairly well in the form of a dual superconductor. This idea was suggested by 't~Hooft and Mandelstam in Ref.~\cite{Mandelstam}. As of now, there is no any strict analytic proof of this assumption. Numerous lattice calculations provide support for that hypothesis (see, for example, Refs.~\cite{Bali:1994de,Chernodub:1997ay}). In Refs.~\cite{DiGiacomo:1999yas,DiGiacomo:1999fb}, the dual superconductivity of the ground state of SU(2) gauge theory is studied in connection with confinement. The investigation of the confining properties of the QCD vacuum of dynamical quarks has been carried out in Ref.~\cite{Bornyakov:2003vx}.
The dual superconductivity model for confinement in QCD is discussed in detail in Ref.~\cite{Ripka:2003vv}. The progress in a gauge-invariant understanding of quark confinement based on the dual superconductivity in Yang-Mills theory is discussed in Ref.~\cite{Kondo:2014sta}.

In the present paper we study configurations consisting of  a quantum monopole (or a dyon) and virtual quarks embedded in a condensate supported by the SU(3)/(SU(2)~$\times$~U(1)) gauge fields. The condensate creates a bag within which a quantum monopole (or a dyon) is located. This means that we are dealing with the Meissner effect: the condensate [supported by the gauge coset fields $A^m_\mu \in$ SU(3) / (SU(2)~$\times$~U(1))] expels the quantum dyon/monopole created by the  $A^a_\mu \in$ SU(2)~$\times$~U(1)~$\subset$~SU(3) fields.
To describe such systems, we use the three-equation approximation of Ref.~\cite{Dzhunushaliev:2016svj} when the right-hand sides of these equations contain sources in the form of a spinor (quark) field for which there is the corresponding nonlinear Dirac equation. The three-equation approximation is the practical realization of the nonperturbative quantization  method suggested by W.~Heisenberg in Ref.~\cite{heis}. He used this approach to describe the  properties of an electron employing some fundamental equation, which he suggested to be  the nonlinear Dirac equation.  Within this direction, the energy spectrum of spherically symmetric configurations has been obtained in Refs.~\cite{Finkelstein:1951zz,Finkelstein:1956}, whose analysis showed the presence of a mass gap. Apparently, the mass gap was first obtained in the theoretical computations of Refs.~\cite{Finkelstein:1951zz,Finkelstein:1956}.

The paper is organized as follows. In Sec.~\ref{quasiPrtcQGP} we discuss quasi-particles in a quark-gluon plasma
and show how solutions obtained in the subsequent sections can be related to quasi-particles. In Sec.~\ref{3eqappr} we present the three-equation approximation to describe the system consisting of virtual quarks, color non-Abelian gauge fields, and the condensate. In Sec.~\ref{QMplusQuarks} we choose the stationary ansatz  to solve the equations of Sec.~\ref{3eqappr}, using which the corresponding complete set of equations is written down. For this set, we seek regular solutions describing three special cases where there present either only quarks (a spinball, Sec.~\ref{spinball}), or  a color magnetic field and the condensate (a quantum monopole, Sec.~\ref{QM}), or a color magnetic field, the condensate, and
quarks (a spinball-plus-quantum-monopole system, Sec.~\ref{QMS}). Also, we obtain regular solutions for the general case where one has color electric and magnetic fields, the condensate, and quarks (a spinball-plus-quantum-dyon system, Sec.~\ref{QDS}). In Sec.~\ref{MassGap} we find energy spectra for the spinball (Sec.~\ref{MassGapSB}) and for the spinball-plus-quantum-monopole system (Sec.~\ref{mgsbqm}), using which we show the existence of mass gaps for these spectra. In Sec.~\ref{deconf} we give a qualitative discussion of the deconfinement mechanism for the spinball and for the spinball-plus-quantum-monopole system when the nonperturbative effects are taken into account. In Sec.~\ref{LambdaQCD} we discuss the issue of the appearance  of the constant $\Lambda_{\text{QCD}}$ in QCD from using the nonperturbative approach \`a la Heisenberg. Finally, in Sec.~\ref{concl}, we summarize the obtained results and give a word about possible physical applications of the systems under consideration.

\section{Quasi-particles in a quark-gluon plasma}
\label{quasiPrtcQGP}

Lattice \cite{Karsch:2002wv,Karsch:2000ps,Laursen:1987eb,Koma:2003hv,Bornyakov:2003vx} and analytical investigations \cite{Shuryak:2004tx,Liao:2006ry,Ramamurti:2017fdn,Ramamurti:2018evz,Shuryak:2018ytg}
indicate that the  quark-gluon plasma contains various quasi-particles: monopoles, dyons, binary bound states (quark-quark ($qq$), quark-antiquark ($q \bar q$), gluon-gluon ($gg$), quark-gluon ($qg$), etc.).
Analytical calculations are phenomenological and they do not provide a macroscopical description of such objects.
In Ref.~\cite{Ramamurti:2018evz}, there is the following assessment of the state of this problem for a monopole: ``\ldots we do not have a microscopic description of these monopoles in terms of the gauge fields.''

The purpose of the present paper is to get a microscopic description of possible quasi-particles in a quark-gluon plasma
based on some approximation for an infinite set of Dyson-Schwinger equations for nonperturbative Green functions. Consistent with this,
below we describe possible types of quasi-particles in a quark-gluon plasma considered in the literature, and for some of them we present the resulting characteristics
which can be compared with the characteristics obtained in our investigations.

\subsection{Magnetic monopoles}
\label{quasiPartMnpl}

One of proofs of the existence of magnetic monopoles in a quark-gluon plasma  is the calculation of the magnetic field flux created by such monopoles.
Ref.~\cite{Bornyakov:2003vx} considers the behavior of the magnetic field flux $\Phi(r)$ as a function of distance from the center of a monopole,
\begin{equation}\label{2a-10}
	\Phi(r) = \Phi_0 \exp \left( -\frac{L}{2 \xi} \right)
	\sinh \left( \frac{L - 2 r}{2 \xi} \right),
\end{equation}
where $L$ is the effective length of the box and $\xi$ is the magnetic screening length.
By going to the continuous limit $L \rightarrow \infty$,
we get the following expression for the  magnetic field flux:
\begin{equation}\label{2a-20}
	\Phi(r) = \Phi_0 \exp \left( -\frac{r}{\xi} \right).
\end{equation}
In Sec.~\ref{QM} the asymptotic expression  \eqref{2-b-50} for the gauge field potential will be obtained,
for which the corresponding radial component of color magnetic field is given by the expression
\begin{equation}\label{2a-30}
	H^{1,2,3}_r \approx H_0 \frac{\exp \left( - \frac{r}{l_0} \right)}{r^2},
\end{equation}
where the magnetic screening length $l_0$ is related to parameters of the system. Calculating this field flux through the sphere with the radius  $r$,
one can derive the expression~\eqref{2a-20}.

In Ref.~\cite{Ramamurti:2018evz}, the following indirect proof of the existence of monopoles is given: one calculates a
semiclassical partition function that can be Poisson-rewritten into an identical ``H'' form. It is shown that it can be done for a pure gauge theory.
After that point, it is argued that the resulting partition function can be interpreted as being generated by moving and rotating monopoles.

In Sec.~\ref{QM} we obtain a solution that describes a ``quantum monopole'' with an exponentially decaying radial magnetic field that is needed in order to explain
the lattice results.

\subsection{Binary bound states}

Another possible quasi-particles in a quark-gluon plasma are binary bound states
which describe states of two particles: $qq$, $qg$, $gg$, etc. In Ref. \cite{Shuryak:2004tx}, it is noted that ``\ldots these bound states are very important for the thermodynamics of the QGP.''
It is pointed out in that paper that in order to describe such objects approximately, one can use either the Klein-Gordon equation,
or the Dirac equation, or the Proca equation. The essence of the suggested approach consists in that these equations are employed to describe two particles, interacting so that they create a coupled pair.
To describe the coupling potential, one uses lattice calculations,
based on which the analytical approximate expression for the potential is suggested.

In Sec.~\ref{spinball} we obtain a solution describing two quarks in a virtual state. This means that the quantum average of the corresponding spinor is zero,
\begin{equation}\label{2a-40}
	\left\langle \hat \psi \right\rangle = 0,
\end{equation}
but the dispersion of such quantum state is nonzero. Physically, this means that the obtained solution describes a quantum object for which the average of field is zero but there exist
quantum fluctuations whose dispersion differs from zero in some region. We assume that this solution microscopically and approximately describes the binary bound state $qq$
where we neglect the distance between quarks and for which the orbital quantum number is zero.

\section{Three-equation approximation with quarks}
\label{3eqappr}

Three-equation approximation has the following form (for a detailed discussion, see Ref.~\cite{Dzhunushaliev:2016svj}):
\begin{eqnarray}
	D_\nu F^{a \mu \nu} - \left[
	\left( m^2 \right)^{ab \mu \nu} -
	\left( \mu^2 \right)^{ab \mu \nu}
	\right] A^b_\nu &=&
	\frac{g \hbar c}{2} \left\langle
	\hat{\bar \psi} \gamma^\mu \lambda^a \hat \psi
	\right\rangle ,
\label{1-10}\\
 	\left\langle
 		\hat A^m_\mu(y) D_\nu \hat F^{m \mu \nu}(x)
 	\right\rangle &=& \frac{g \hbar c}{2} \left\langle
 			\hat A^m_\mu(y) \hat{\bar \psi}(x) \gamma^\mu \lambda^m \hat \psi(x)
 		\right\rangle ,
\label{1-20}\\
	\left\langle \hat \psi_{i \alpha}(y) \left\lbrace
		i \hbar \gamma^\mu \left[
		\partial_\mu \hat \psi (x) - i \frac{g}{2} \lambda^B
		\hat A^B_\mu (x) \hat \psi (x)
		\right] -m_q c \hat \psi (x) \right\rbrace_{j \beta}
	\right\rangle &=& 0,
\label{1-30}
\end{eqnarray}
where $i,j = \left\lbrace \text{red, green, blue}\right\rbrace$ are the color indices of quarks; $\alpha, \beta$ are the spinor indices of quarks; $g$ is the coupling constant; $m_q$ is the quark mass; $D_\nu$ is the gauge derivative of the subgroup SU(2)~$\times$~U(1); $\left( m^2 \right)^{ab \mu\nu}$,
$\left( \mu^2 \right)^{ab \mu \nu}$, and $\left( m^2_\phi \right)^{ab \mu \nu}$ [see below in Eq.~\eqref{1-52}] are quantum corrections coming from the dispersions of the operators $\widehat{\delta A}^{a \mu}$ and $\hat A^{m \mu}$,
\begin{eqnarray}
	\hat A^{a \mu} &=& \left\langle \hat A^{a \mu} \right\rangle +
	i \widehat{\delta A}^{a \mu} ,
\label{1-40}\\
	\left\langle \hat A^{m \mu} \right\rangle &=& 0 ,
\label{1-50}
\end{eqnarray}
where $\left\langle \ldots \right\rangle$ denotes the quantum average over some nonperturbative quantum state; $a$ is the index of the
$SU(2) \times U(1) \subset SU(3)$ subgroup; $\lambda^B$ are the Gell-Mann matrices; $B = 1,2, \ldots , 8$ is the index of the  SU(3) group. The dispersions of the quantities in
Eqs.~\eqref{1-40} and \eqref{1-50} are defined in Sec.~\ref{LambdaQCD} by the formulae \eqref{4-20} and \eqref{4-30}. The notion of the nonperturbative quantum state is discussed in detail in Ref.~\cite{Dzhunushaliev:2017fqa}.

The left-hand side of Eq.~\eqref{1-20} is derived in analogy with obtaining this equation with zero right-hand side in Ref.~\cite{Dzhunushaliev:2016svj}:
\begin{equation}
 	\left\langle
 		\hat A^m_\mu(y) D_\nu \hat F^{m \mu \nu}(x)
 	\right\rangle = \phi(y) \left[
		\Box \phi - \left( m^2_\phi \right)^{ab \mu \nu} A^a_\nu A^b_\mu \phi -
		\lambda \phi \left( M^2 - \phi^2 \right)
	\right] .
\label{1-52}
\end{equation}
Using the method developed in~\cite{Dzhunushaliev:2016svj}, one then can obtain Eq.~\eqref{1-20}.

To estimate the right-hand sides of Eqs.~\eqref{1-10}-\eqref{1-30}, we use approximations according to which
\begin{eqnarray}
	\left\langle \hat \psi \right\rangle &=& 0 ,
\label{1-60}\\
	\left\langle
		\hat{\bar \psi} \gamma^\mu \lambda^a \hat \psi
	\right\rangle &=& \bar \zeta \gamma^\mu \lambda^a \zeta ,
\label{1-70}\\
	\left\langle
		\hat{\bar \psi}_{i \alpha}(y) \lambda^m_{j k} \left(
			\gamma^\mu
		\right)_{\beta \gamma} \hat A^m_\mu(y)
		\hat \psi_{\gamma k}
	\right\rangle &=& ?
\label{1-95}
\end{eqnarray}
Here $\zeta$ is the spinor describing the dispersion \eqref{1-70} of the quantum field $\hat \psi$ which has a zero average value \eqref{1-60}. In this sense we consider virtual quarks for which the quantum average of $\hat \psi$ is equal to zero.

The relation \eqref{1-95}, using Eq.~\eqref{1-50}, is of most interest. Let us consider the Green function \eqref{1-95} in more detail:
\begin{equation}
	G_{\alpha \beta i j}(y,y,x) =
	\left\langle
		\hat{\bar \psi}_{\alpha i}(y) \left( \lambda^m \right)_{jk}
		\left( \gamma^\mu \right)_{\beta \gamma}
		\hat A^m_\mu(y)
		\hat \psi_{\gamma k}(x)
	\right\rangle.
\label{1-105}
\end{equation}
Here $\alpha, \beta, \gamma$ are the spinor indices and $i,j, k = 1 ,2, 3$ (or
$i,j,k = \left\lbrace \text{red, green, blue}\right\rbrace$) are the quark indices. One of our main purposes is to obtain an approximate expression for the Green function  $G_{\alpha \beta i j}$. This function has the indices  $\alpha, \beta$ and $i, j$, and,  hence, our approximation must also have the same indices. For this purpose, we will use the dispersion of the spinor field  $\zeta_{\alpha}$ and the dispersion of the coset gauge field $\phi$. One can construct several different variants to approximate the Green function $G_{\alpha \beta i j}$ from \eqref{1-105}. The following approximation will be the simplest one:
\begin{eqnarray}
	G_{\alpha \beta i j}(y,x,x) &\approx&
	\Lambda_1 \left[
		\bar \zeta_{\alpha i}(y) \zeta_{\beta j}(x)
	\right] \left[
		\bar \zeta_{k \gamma}(x) \zeta_{k \gamma}(x)
	\right]
	\phi(x) = \Lambda_1 \left[
		\bar \zeta_{\alpha i}(y) \zeta_{\beta j}(x)
	\right] \left( \bar \zeta \zeta \right) \phi(x) ,
\label{1-102}\\
	G_{\alpha \alpha i i}(x,y,x) &\approx&
	\Lambda_2 \left[
		\bar \zeta_{\alpha i}(y) \zeta_{\alpha i}(x)
	\right] \left[
		\bar \zeta_{k \gamma}(x) \zeta_{k \gamma}(x)
	\right]
	\phi(y) =
	\Lambda_2 \left( \bar \zeta \zeta \right)^2 \phi(y).
\label{1-104}
\end{eqnarray}
Here $\Lambda_{1,2}$ are some numerical coefficients. The approximation
\eqref{1-102} is substituted into  Eq.~\eqref{1-30}; cancelling
$\bar \zeta_{\alpha i}(y)$, we obtain the nonlinear Dirac equation \eqref{1-140} for $\zeta$. The approximation \eqref{1-104} is substituted into the right-hand side of Eq.~\eqref{1-20}; cancelling $\phi(y)$, we obtain Eq.~\eqref{1-130} for the condensate $\phi$. Physically, this approximation means that the Green function $G_{\alpha \beta i j}$, describing the correlation between quantum fluctuations of the fields $\hat A^m_\mu$ and $\hat \psi$, depends on the dispersions of quantum fluctuations of the field $\hat A^m_\mu$ and the quark field $\hat \psi$.

Hence, using the above approximations, Eqs.~\eqref{1-10}-\eqref{1-30} take the final form
\begin{eqnarray}
	D_\nu F^{a \mu \nu} - \left[
		\left( m^2 \right)^{ab \mu \nu} -
		\left( \mu^2 \right)^{ab \mu \nu}
	\right] A^b_\nu &=& \frac{g \hbar c}{2} \left(
		\bar \zeta \gamma^\mu \lambda^a \zeta
	\right) ,
\label{1-120}\\
	\Box \phi - \left( m^2_\phi \right)^{ab \mu \nu} A^a_\nu A^b_\mu \phi -
	\lambda \phi \left( M^2 - \phi^2 \right) &=&
	\Lambda_2 \frac{g \hbar c}{2} \left(
		\bar \zeta \zeta
	\right)^2 ,
\label{1-130}\\
	i \hbar \gamma^\mu \left(
		\partial_\mu \zeta - i \frac{g}{2} \lambda^a
		A^a_\mu \zeta
	\right)
	+ \Lambda_1 \frac{g \hbar}{2} \phi \zeta
	\left(
		\bar \zeta \zeta
	\right) - m_q c\, \zeta &=& 0.
\label{1-140}
\end{eqnarray}

\section{Quantum dyon plus virtual quarks}
\label{QMplusQuarks}

We seek a solution of Eqs.~\eqref{1-120}-\eqref{1-140} in the following form:
\begin{eqnarray}
	A^a_i &=& \frac{f(r) - 1}{g r^2} \epsilon_{i a j} x^j , \quad
	i = x,y,z  \text{ (in cartesian coordinates)},
\label{2-10}\\
	A^a_i &=& \frac{1}{g} \left[ 1 - f(r) \right]
	\begin{pmatrix}
		 0 & \phantom{-}\sin \varphi &  \sin \theta \cos \theta \cos \varphi \\
		 0 & -\cos \varphi &  \sin \theta \cos \theta \sin \varphi \\
		 0 & 0 & - \sin^2 \theta
	\end{pmatrix} , \quad
	i = r, \theta, \varphi  \text{ (in polar coordinates)},
\label{2-12}\\
	A^a_t &=& 0 ,
\label{2-13}\\
	A^8_t &=& \frac{\chi(r)}{g} , \quad A^8_i = 0 ,
\label{2-15}\\
	\phi &=& \frac{\xi(r)}{g},
\label{2-20}\\
	\zeta^T &=& \frac{e^{-i \frac{E t}{\hbar}}}{g r \sqrt{2}} \begin{Bmatrix}
		\begin{pmatrix}
			0 \\ - u \\ 0 \\
		\end{pmatrix},
		\begin{pmatrix}
			u \\ 0 \\ 0 \\
		\end{pmatrix},
		\begin{pmatrix}
			i v \sin \theta e^{- i \varphi} \\ - i v \cos \theta \\ 0 \\
		\end{pmatrix},
		\begin{pmatrix}
			- i v \cos \theta \\ - i v \sin \theta e^{i \varphi} \\ 0 \\
		\end{pmatrix}
	\end{Bmatrix}.
\label{2-30}
\end{eqnarray}
Here $\epsilon_{i a j}$ is the completely antisymmetric Levi-Civita symbol; $a=1,2,3$;  $i,j = 1,2,3$ are the spacetime indices; the functions $u$ and $v$ depend on the radial coordinate $r$ only; the ansatz~\eqref{2-30} is taken from Refs.~\cite{Li:1982gf,Li:1985gf}. After substituting the expressions \eqref{2-10}-\eqref{2-30} into Eqs.~\eqref{1-120}-\eqref{1-140},
we obtain the equations
\begin{eqnarray}
	- f^{\prime \prime} + \frac{f \left( f^2 - 1 \right) }{x^2} -
	m^2 \left( 1 - f \right) \tilde \xi^2 +
	\tilde g^2\,\frac{\tilde u \tilde v}{x} &=& - \tilde{\mu}^2 \left( 1 - f \right) ,
\label{2-40}\\
	\tilde \chi'' + \frac{2}{x} \tilde \chi' &=& \frac{1}{2\sqrt{3}}\,\tilde g^2\,
	\frac{\tilde u^2 + \tilde v^2}{x^2} ,
\label{2-45}\\
	\tilde \xi^{\prime \prime} + \frac{2}{x} \tilde \xi^\prime &=&
	\tilde \xi \left[
		\frac{\left( 1 - f \right)^2}{2 x^2}  +
		\tilde{\lambda} \left(
			\tilde \xi^2 - \tilde{M}^2
		\right)
	\right] -
	\frac{\tilde g^2\tilde\Lambda}{8}
	\frac{\left( \tilde u^2 - \tilde  v^2 \right)^2}{x^4} ,
\label{2-50}\\
	\tilde v' + \frac{f \tilde v}{x} &=& \tilde u \left(
		- \tilde m_q + \tilde E +
		m^2 \tilde \Lambda   \frac{\tilde u^2 - \tilde v^2}{x^2} \tilde \xi +
		\frac{\tilde \chi}{2 \sqrt{3}}
	\right) ,
\label{2-60}\\
	\tilde u' - \frac{f \tilde u}{x} &=& \tilde v \left(
		- \tilde m_q - \tilde E +
				m^2 \tilde \Lambda   \frac{\tilde u^2 - \tilde v^2}{x^2} \tilde \xi-
		\frac{\tilde \chi}{2 \sqrt{3}}
	\right).
\label{2-70}
\end{eqnarray}
In these equations, the following dimensionless variables are used:
$\tilde g^2 = g^2 \hbar c$ is the dimensionless coupling constant for the  SU(3) gauge field; $x = r/r_0$, where $r_0$ is a constant corresponding to the characteristic size of the system under consideration (in Sec.~\ref{LambdaQCD} we will show that this parameter is related to the constant $\Lambda_{\text{QCD}}$);
$\tilde u=\sqrt{r_0}u/g$, $\tilde v=\sqrt{r_0}v/g,\tilde \mu =r_0 \mu, \tilde \xi =r_0 \xi,
\tilde \chi =r_0 \chi, \tilde M=g r_0 M, \tilde \lambda=\lambda/g^2, \tilde m_q=c r_0 m_q/\hbar,
\tilde E=r_0 E/(\hbar c), \tilde \Lambda=\Lambda/r_0^3$.

In order that these equations would be Euler-Lagrange equations it is necessary to choose the following values of the dimensionless constants:
$\Lambda_1 = 2 m^2 \Lambda, \Lambda_2 = \Lambda/4$. Then the dimensionless effective Lagrangian for this set of equations is as follows:
\begin{equation}
\begin{split}
	\mathcal{\tilde{L}_{\text{eff}}} \equiv & \mathcal{L_{\text{eff}}}/\left(\hbar c/r_0^4\right) =
 	\frac{1}{\tilde g^2} \left\lbrace \frac{\tilde \chi^{\prime 2}}{2} -
	\left[
		\frac{{f'}^2}{ x^2} +
		\frac{\left( f^2 - 1 \right)^2}{2 x^4} -
		\tilde \mu^2 \frac{\left( f - 1 \right)^2}{x^2}
	\right] -
	2 m^2 \left[
		\tilde \xi^{\prime 2} +
		\frac{\left( f - 1 \right)^2}{2 x^2} \tilde \xi^2 +
		\frac{\tilde \lambda}{2}\left(
		\tilde \xi^2 - \tilde M^2
		\right)^2
	\right]
	\right\rbrace
\\
	&	
	+ \frac{1}{ x^2} \left[
		- \tilde u \tilde v' + \tilde u' \tilde v -
		2 f \frac{\tilde u \tilde v}{x} +
		m^2 \frac{\tilde\Lambda}{2} \tilde \xi
		\frac{\left(\tilde u^2 - \tilde v^2 \right)^2}{x^2} -
		\tilde m_q \left(\tilde  u^2 - \tilde v^2 \right) +
		\tilde E \left(\tilde u^2 + \tilde  v^2 \right) +
		\frac{\tilde \chi }{2 \sqrt{3}} \left(\tilde  u^2 + \tilde  v^2 \right)
		\right] .
\label{2-80}
\end{split}
\end{equation}
Next, by definition, the energy density of the Dirac field is
\begin{equation}
	\epsilon_D =i\hbar \bar\zeta \gamma^0 \dot\zeta - L_{D\text{eff}},
\label{2-90}
\end{equation}
where the dot denotes differentiation with respect to time. The Lagrangian of the Dirac field $L_{D\text{eff}}$ appearing here is given by the expression from  \eqref{2-80},
\begin{equation}
\begin{split}
	\tilde L_{D\text{eff}} = &
	\frac{1}{ x^2} \left[
		- \tilde u \tilde v' + \tilde u' \tilde v -
		2 f \frac{\tilde u \tilde v}{x} +
		m^2 \frac{\tilde\Lambda}{2} \tilde \xi
		\frac{\left(\tilde u^2 - \tilde v^2 \right)^2}{x^2} -
		\tilde m_q \left(\tilde  u^2 - \tilde v^2 \right) +
		\tilde E \left(\tilde u^2 + \tilde  v^2 \right) +
		\frac{\tilde \chi }{2 \sqrt{3}} \left(\tilde  u^2 + \tilde  v^2 \right)
		\right]
\\
	&
	=-m^2 \frac{\tilde\Lambda}{2 } \tilde \xi
	\frac{\left(\tilde u^2 - \tilde v^2 \right)^2}{x^4},
\end{split}
\label{2-100}
\end{equation}
which is obtained using Eqs.~\eqref{2-60} and \eqref{2-70}. Then, using the ansatz \eqref{2-30}, the energy density of the spinor field \eqref{2-90} can be found in the form
\begin{equation}
	\tilde \epsilon_D = \tilde E \frac{\tilde u^2 + \tilde v^2}{x^2} +
	m^2 \frac{\tilde\Lambda}{2 } \tilde \xi
	\frac{\left(\tilde u^2 - \tilde v^2 \right)^2}{x^4} .
\label{2-110}
\end{equation}
As a result, we get the following total energy density of the system:
\begin{equation}
\begin{split}
	\tilde \epsilon = &
 	\frac{1}{\tilde g^2} \left\lbrace \frac{\tilde \chi^{\prime 2}}{2} +
	\left[
		\frac{{f'}^2}{ x^2} +
		\frac{\left( f^2 - 1 \right)^2}{2 x^4} -
		\tilde \mu^2 \frac{\left( f - 1 \right)^2}{x^2}
	\right] +
	2 m^2 \left[
		\tilde \xi^{\prime 2} +
		\frac{\left( f - 1 \right)^2}{2 x^2} \tilde \xi^2 +
		\frac{\tilde \lambda}{2}\left(
		\tilde \xi^2 - \tilde M^2
		\right)^2
	\right]
	\right\rbrace
\\
	&	
	+\tilde E \frac{\tilde u^2 + \tilde v^2}{x^2} +
	m^2 \frac{\tilde\Lambda}{2 } \tilde \xi
	\frac{\left(\tilde u^2 - \tilde v^2 \right)^2}{x^4}  +
		\tilde \epsilon_\infty,
\label{2-120}
\end{split}
\end{equation}
where the arbitrary constant $\tilde \epsilon_\infty$ corresponding to the energy density at infinity has been also introduced.

\subsection{Spinball}
\label{spinball}

Solving the complete set of equations \eqref{2-40}-\eqref{2-70} runs into great difficulty, and hence we start from considering the simplest problem
without the color fields. In doing so, we assume that two virtual quarks are located at the center. Such a consideration enables one to simplify
a mathematical study of the system: it allows using the ansatz employed in describing the quantum state of an electron in a hydrogen atom.
Otherwise, it would be necessary to solve a problem analogous to that which occurs in describing the quantum state of two separated electrons
in a  helium atom.

\begin{figure}[t]
	\begin{minipage}[t]{.45\linewidth}
		\begin{center}
			\includegraphics[width=1\linewidth]{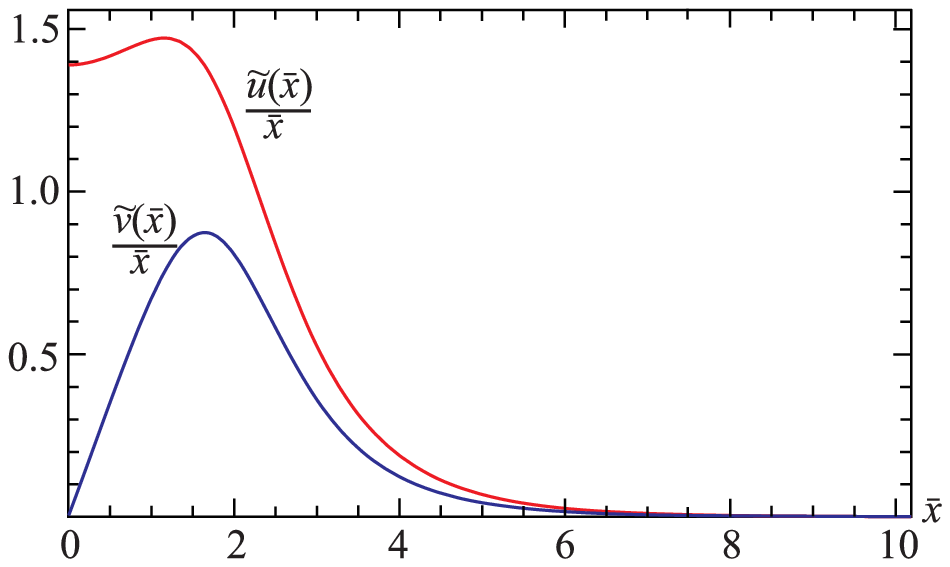}
		\end{center}
\vspace{-0.5cm}
		\caption{The typical behavior of the solutions of Eqs.~\eqref{2-a-20} and \eqref{2-a-30}
describing the spinball for	$\bar \Lambda = 1$ and $\bar E = 0.6$. The corresponding eigenvalue is	$\tilde u_1 \approx 1.389621$.
		}
		\label{uv}
	\end{minipage}\hfill
\begin{minipage}[t]{.45\linewidth}
		\begin{center}
			\includegraphics[width=1\linewidth]{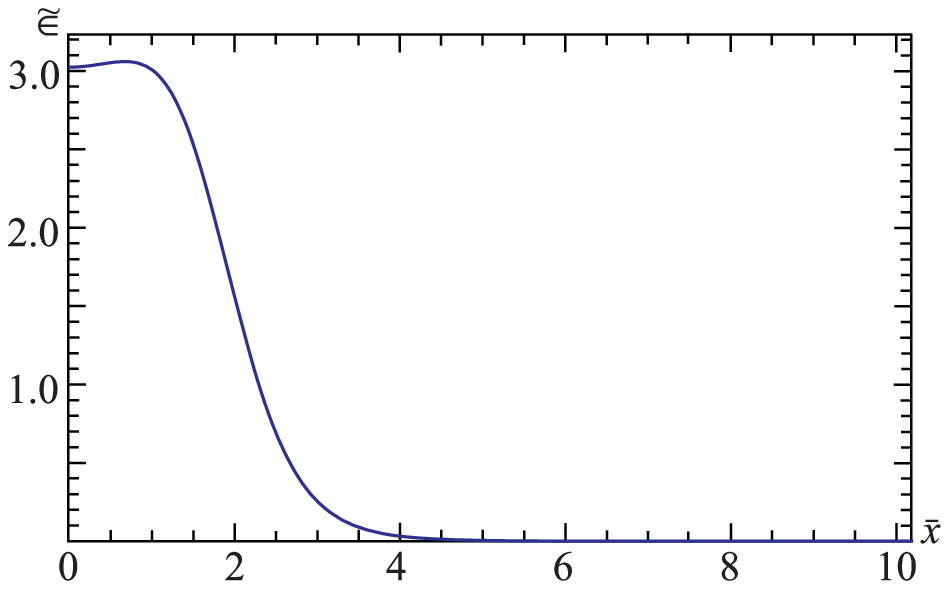}
		\end{center}
\vspace{-0.5cm}
		\caption{The graph of the dimensionless energy density $\tilde \epsilon(\bar x)/\tilde m_q^3$ from \eqref{2-a-100} for the spinball.
		 		}
		\label{EnDensitySB}
	\end{minipage}
\end{figure}

For such a system, there are only Eqs.~\eqref{2-60} and \eqref{2-70}, in which we must take  $f(x) = 1$ and $\tilde\xi =\tilde M$. The resulting equations are then
\begin{eqnarray}
	\tilde v' + \frac{\tilde v}{\bar x} &=& \tilde u \left(
		- 1 + \bar E + \bar \Lambda \frac{\tilde u^2 - \tilde v^2}{\bar x^2}
	\right) ,
\label{2-a-20}\\
	\tilde u' - \frac{\tilde u}{\bar x} &=& \tilde v \left(
		- 1 - \bar E + \bar \Lambda \frac{\tilde u^2 - \tilde v^2}{\bar x^2}
	\right).
\label{2-a-30}
\end{eqnarray}
Here, we have introduced new dimensionless variables
$\bar x = \tilde m_q x$, $\bar E = \tilde E/\tilde m_q$, and $\bar \Lambda= m^2 \tilde m_q \tilde M \tilde\Lambda $.
Regular solutions to Eqs.~\eqref{2-a-20} and \eqref{2-a-30} can be used in describing a spinball -- a finite-size system consisting of two virtual quarks with opposite spins and a zero distance between them. In the neighborhood of the center of such a spinball, solutions are sought in the form
 \begin{eqnarray}
	\tilde u &=& \tilde u_1 \bar x + \frac{\tilde u_3}{3!} \bar x^3 + \ldots ,
\label{2-a-50}\\
	\tilde v &=& \frac{\tilde v_2}{2} \bar x^2 + \frac{\tilde v_4}{4!} \bar x^4 + \ldots , \quad \text{where} \quad
	\tilde v_2 = \frac{2}{3} \tilde u_1 \left(
		- 1+ \bar E +  \bar \Lambda \tilde u_1^2
	\right) .
\label{2-a-60}
\end{eqnarray}
Eqs.~\eqref{2-a-20} and \eqref{2-a-30} are solved numerically as a nonlinear problem for the eigenvalue $\bar E$ (or $\tilde u_1$)  and the eigenfunctions $\tilde v(\bar x)$ and $\tilde u(\bar x)$. An example of a typical solution is shown in Fig.~\ref{uv}, with the asymptotic behavior
 \begin{equation}
	\tilde u \approx \tilde v_\infty e^{- \bar x
	\sqrt{1 - \bar E^2}} , \quad \tilde v \approx \tilde u_\infty e^{- \bar x
	\sqrt{1 - \bar E^2}},
\label{2-a-80}
\end{equation}
where $\tilde u_\infty$ and $\tilde v_\infty$ are integration constants.

The dimensionless energy density of the spinball under consideration can be found from Eq.~\eqref{2-120} in the form
\begin{equation}
	\tilde \epsilon(\bar x)/\tilde m_q^3 = \bar E \frac{\tilde u^2 + \tilde v^2}{\bar x^2} +
			\frac{\bar \Lambda}{2 }
			\frac{\left(\tilde u^2 - \tilde v^2 \right)^2}{\bar x^4} .
\label{2-a-100}
\end{equation}
The corresponding graph is shown in Fig.~\ref{EnDensitySB} for the values of the parameters of the system given in caption to Fig.~\ref{uv}.
Note that here and in Secs.~\ref{QM}, \ref{QMS}, and \ref{QDS} we set $\tilde \epsilon_\infty = 0$.	

\subsection{Quantum monopole}
\label{QM}

\begin{figure}[t]
	\begin{minipage}[t]{.45\linewidth}
		\begin{center}
			\includegraphics[width=1\linewidth]{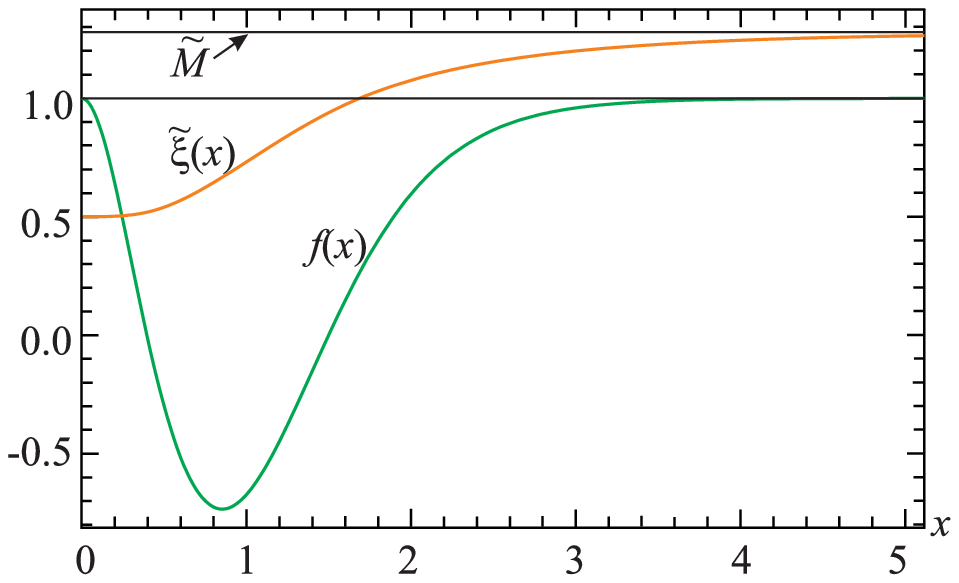}
		\end{center}
\vspace{-0.5cm}
		\caption{The typical behavior of the functions $f(x)$ and $\tilde\xi(x)$ of the solution describing the
 quantum monopole at $f_2 = -20$,
		  $m = 3$, $\tilde\lambda = 0.1$, and $\tilde\xi(0) = 0.5$. The corresponding eigenvalues are
		  $\tilde \mu = 2.68238579$ and
		  $\tilde M = 1.2786$.
		 		}
		\label{fphi}
	\end{minipage}\hfill
\begin{minipage}[t]{.45\linewidth}
		\begin{center}
			\includegraphics[width=1\linewidth]{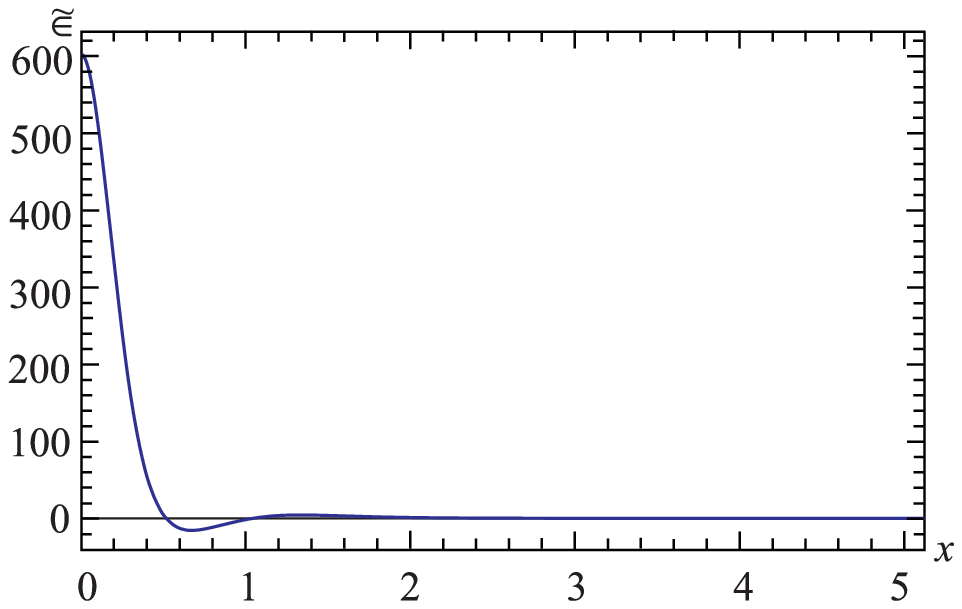}
		\end{center}
\vspace{-0.5cm}
		\caption{The graph of the dimensionless energy density \eqref{2-b-70} of the quantum monopole.
		 		}
		\label{EnDensityQM}
	\end{minipage}
\end{figure}

In this section we consider one more simplified configuration -- the quantum monopole \cite{Dzhunushaliev:2017rin}. In doing so, we will use Eqs.~\eqref{2-40} and \eqref{2-50} with the color gauge field (i.e., when $f(x) \neq 1$) and without quarks (i.e., when $\tilde v(x) = \tilde u(x) = 0$):
\begin{eqnarray}
	- f^{\prime \prime} + \frac{f \left( f^2 - 1 \right) }{x^2} -
	m^2 \left( 1 - f \right) \tilde\xi^2 &=& - \tilde{\mu}^2 \left( 1 - f \right) ,
\label{2-b-10}\\
	\tilde\xi^{\prime \prime} + \frac{2}{x} \tilde\xi^\prime &=&
	\tilde\xi \left[
	\frac{\left( 1 - f \right)^2}{2 x^2}  +
	\tilde{\lambda} \left(
	\tilde\xi^2 - \tilde{M}^2
	\right)
	\right] .
\label{2-b-20}
\end{eqnarray}
We seek solutions in the neighborhood of the center in the form
\begin{eqnarray}
	f &=& 1 + \frac{f_2}{2} x^2 + \ldots ,
\label{2-b-30}\\
	\tilde\xi &=& \tilde\xi_0 + \frac{\tilde\xi_2}{2} x^2 + \ldots ,
\label{2-b-40}
\end{eqnarray}
where the expansion coefficient $f_2$ is arbitrary and
$\tilde\xi_2 = \tilde\lambda \tilde\xi_0 (\tilde\xi_0^2 - \tilde{M}^2)/3$.
Using these boundary conditions, Eqs.~\eqref{2-b-10} and \eqref{2-b-20} are solved as a nonlinear problem for the eigenvalues $\tilde M, \tilde \mu$ and the eigenfunctions $\tilde \xi(x), f(x)$. The results of numerical calculations are given in Fig.~\ref{fphi}. The corresponding asymptotic behavior of these solutions is as follows:
 \begin{equation}
	f(x) \approx 1 - f_\infty e^{- x \sqrt{m^2 \tilde{M}^2 - \tilde\mu^2}} , \quad
	\tilde\xi (x) \approx \tilde M -
	\tilde\xi_\infty \frac{e^{- x \sqrt{2 \tilde\lambda \tilde{M}^2}}}{x},
\label{2-b-50}
\end{equation}
where $\tilde\xi_\infty, f_\infty$ are integration constants.
The obtained solutions describe a quantum monopole placed in the coset condensate, and such a quantum monopole is expelled from the condensate
due to the Meissner effect.

The radial color magnetic field is defined as follows:
\begin{equation}\label{2-b-53}
		H^a_r = \frac{1 - f^2}{g r^2}.
\end{equation}
Its asymptotic behavior is
\begin{equation}\label{2-b-56}
	H^a_r(r) \approx \frac{2}{g}
	\frac{e^{-\frac{r}{r_0} \sqrt{m^2 \tilde M^2 - \tilde \mu^2}}}{r^2}.
\end{equation}
Calculating the flux of this field and comparing its with the expression  \eqref{2a-20},
it is seen that the magnetic screening length $\xi$ is related to the microscopic parameters of the system
$m, \tilde M, \tilde \mu$ as follows:
\begin{equation}\label{2-b-59}
	\xi = \frac{\Lambda_{\text{QCD}}}{\sqrt{m^2 \tilde{M}^2 - \tilde\mu^2}}.
\end{equation}
Here we have identified  the undetermined parameter  $r_0$ with $\Lambda_{\text{QCD}}$
 (as it will be explained in Sec.~\ref{LambdaQCD}) .

The dimensionless energy density of the quantum monopole can be found from Eq.~\eqref{2-120} in the form
\begin{equation}
	\tilde \epsilon = \frac{1}{\tilde g^2} \left\lbrace
	\left[
		\frac{f'^2}{ x^2} +
		\frac{\left( f^2 - 1 \right)^2}{2 x^4} -
		\mu^2 \frac{\left( f - 1 \right)^2}{x^2}
	\right] +
	2 m^2 \left[
		\tilde\xi'^2 +
		\frac{\left( f - 1 \right)^2}{2 x^2} \tilde\xi^2 +
		\frac{\tilde\lambda}{2}\left(
		\tilde\xi^2 - \tilde M^2
		\right)^2
	\right]
	\right\rbrace .
\label{2-b-70}
\end{equation}
The corresponding graph is shown in Fig.~\ref{EnDensityQM} for the values of the parameters of the system given in caption to Fig.~\ref{fphi}.

Consider now the physical meaning of the parameter $f_2$. Using Eq.~\eqref{2-b-30}, for the radial magnetic field $H^a_r$ in the neighbourhood of the origin,
one can find
\begin{equation}
	\tilde H^{1,2,3}_r = \frac{1 - f^2}{x^2} \approx -f_2.
\end{equation}
Hence $f_2$ gives the magnitude of the magnetic field at the center of the quantum monopole.

\subsection{Spinball plus a quantum monopole}
\label{QMS}

\begin{figure}[t]
	\begin{minipage}[t]{.45\linewidth}
		\begin{center}
			\includegraphics[width=1\linewidth]{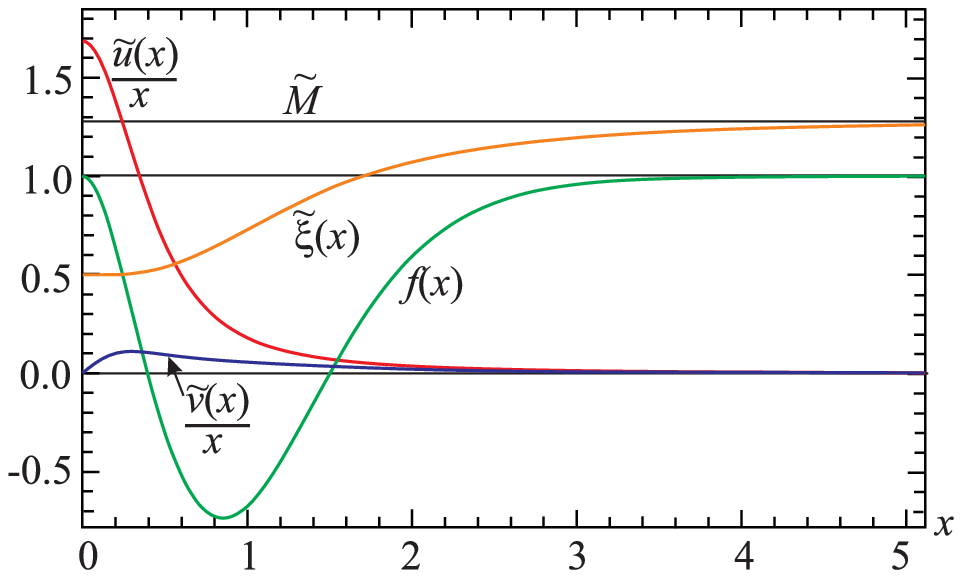}
		\end{center}
\vspace{-0.5cm}
		\caption{The graphs of the functions $\tilde v(x), \tilde u(x), f(x)$, and $\tilde \xi(x)$ of the solution describing the spinball-plus-quantum-monopole system
  at $\tilde \lambda = 0.1$,
			$\tilde \Lambda = 1/9$, $\tilde m_q = 1$, $\tilde E = 0.8$, $m = 3$, $\tilde g=1$,
			$\tilde \xi_0 = 0.5$, and $f_2 = - 20$. The corresponding
			eigenvalues are $\tilde \mu \approx 2.6727842881$,
			$\tilde M \approx 1.2756$, and $\tilde u_1 \approx 1.68655$.
		}
		\label{fuvphi}
	\end{minipage}\hfill
	\begin{minipage}[t]{.45\linewidth}
		\begin{center}
			\includegraphics[width=1\linewidth]{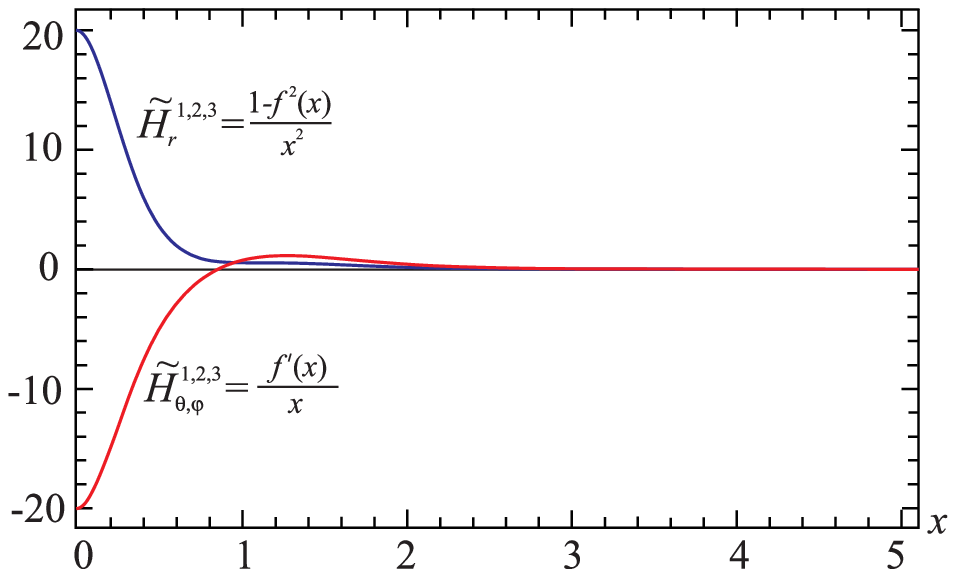}
		\end{center}
\vspace{-0.5cm}
		\caption{	
			The profiles of the magnetic fields $\tilde H^{1,2,3}_r$ and
			$\tilde H^{1,2,3}_{\theta, \varphi}$ for the spinball-plus-quantum-monopole configuration.
		}
		\label{SBQMfields}
	\end{minipage}
\end{figure}

In this section we consider a solution of the set of equations
\begin{eqnarray}
- f^{\prime \prime} + \frac{f \left( f^2 - 1 \right) }{x^2} -
	m^2 \left( 1 - f \right) \tilde \xi^2 +
	\tilde g^2\,\frac{\tilde u \tilde v}{x} &=& - \tilde{\mu}^2 \left( 1 - f \right) ,
\label{3-c-10}\\
	\tilde \xi^{\prime \prime} + \frac{2}{x} \tilde \xi^\prime &=&
	\tilde \xi \left[
		\frac{\left( 1 - f \right)^2}{2 x^2}  +
		\tilde{\lambda} \left(
			\tilde \xi^2 - \tilde{M}^2
		\right)
	\right] -
	\frac{\tilde g^2\tilde\Lambda}{8}
	\frac{\left( \tilde u^2 - \tilde  v^2 \right)^2}{x^4} ,
\label{3-c-20}\\
		\tilde v' + \frac{f \tilde v}{x} &=& \tilde u \left(
		- \tilde m_q + \tilde E +
		m^2 \tilde \Lambda   \frac{\tilde u^2 - \tilde v^2}{x^2} \tilde \xi
	\right) ,
\label{3-c-30}\\
	\tilde u' - \frac{f \tilde u}{x} &=& \tilde v \left(
		- \tilde m_q - \tilde E +
				m^2 \tilde \Lambda   \frac{\tilde u^2 - \tilde v^2}{x^2} \tilde \xi
	\right).
\label{3-c-40}
\end{eqnarray}
These equations describe a system consisting of a quantum monopole, two virtual quarks, an extra color magnetic field created by them, and the condensate.
Because of the Meissner effect, the magnetic field is expelled by the condensate creating the quantum monopole (hedgehog). The color magnetic field is produced both by the quantum monopole and by the virtual quarks. In such a configuration the virtual quarks are located at an infinitely small distance from each other that, as in Sec.~\ref{spinball}, enables one to use for their modeling the ansatz employed in describing an electron in a hydrogen atom.

As $x \rightarrow 0$, the solution of the above set of equations can be presented in the form of the following expansions:
\begin{eqnarray}
	f &=& 1 + \frac{f_2}{2} x^2 + \ldots ,
\label{3-c-50}\\
	\tilde\xi &=& \tilde\xi_0 + \frac{\tilde\xi_2}{2} x^2 + \ldots ,
\label{3-c-60}\\
	\tilde u &=& \tilde u_1 x + \frac{\tilde u_3}{3!} x^3 + \ldots ,
\label{3-c-70}\\
	\tilde v &=& \frac{\tilde v_2}{2} x^2 + \frac{\tilde v_4}{4!} x^4 + \ldots , \quad \text{where} \quad
	\tilde v_2 = \frac{2}{3} \tilde u_1 \left(
		\tilde E-\tilde m_q  + m^2 \tilde\Lambda \tilde\xi_0  \tilde u_1^2
	\right) .
\label{3-c-80}
\end{eqnarray}
Eqs.~\eqref{3-c-10}-\eqref{3-c-40} are solved numerically as a nonlinear problem for the eigenvalues $\tilde \mu, \tilde M, \tilde E$ and the eigenfunctions $f(x), \tilde\xi(x), \tilde v(x), \tilde u(x)$.
Fig.~\ref{fuvphi} shows the typical behavior of the solutions. Their asymptotic behavior as $x\to \infty$ is
\begin{eqnarray}
	f(x) &\approx& 1 - f_\infty e^{- x \sqrt{m^2 \tilde{M}^2 - \tilde\mu^2}} ,
\quad \tilde\xi (x) \approx \tilde M -
	\tilde\xi_\infty \frac{e^{- x \sqrt{2 \tilde\lambda \tilde{M}^2}}}{x},
\label{3-c-90}\\
		\tilde u &\approx& \tilde u_\infty e^{- x \sqrt{
			\tilde m_q^2 - \tilde E^2
	}} , \quad \tilde v \approx \tilde v_\infty e^{- x \sqrt{
	\tilde m_q^2 - \tilde E^2
	}},
\label{3-c-110}
\end{eqnarray}
where $f_\infty, \tilde\xi_\infty, \tilde v_\infty$,  and $\tilde u_\infty$ are integration constants. The corresponding distributions of different components of the color magnetic field are shown in Fig.~\ref{SBQMfields}.

The dimensionless energy density of the system in question can be obtained from \eqref{2-120} in the form
\begin{equation}
\begin{split}
		\tilde \epsilon = &
	 	\frac{1}{\tilde g^2} \left\lbrace
	\left[
		\frac{{f'}^2}{ x^2} +
		\frac{\left( f^2 - 1 \right)^2}{2 x^4} -
		\tilde \mu^2 \frac{\left( f - 1 \right)^2}{x^2}
	\right] +
	2 m^2 \left[
		\tilde \xi^{\prime 2} +
		\frac{\left( f - 1 \right)^2}{2 x^2} \tilde \xi^2 +
		\frac{\tilde \lambda}{2}\left(
		\tilde \xi^2 - \tilde M^2
		\right)^2
	\right]
	\right\rbrace
\\
	&	
	+\tilde E \frac{\tilde u^2 + \tilde v^2}{x^2} +
	m^2 \frac{\tilde\Lambda}{2 } \tilde \xi
	\frac{\left(\tilde u^2 - \tilde v^2 \right)^2}{x^4}.
\label{3-c-130}
\end{split}
\end{equation}
When one chooses, for instance, $f_2=-20$ and $\tilde E=0.8$, and also uses the same values of the free parameters as those for the system from Sec.~\ref{QM} (see in caption to Fig.~\ref{fphi}), its graph practically coincides with that of the energy density of the quantum monopole shown in Fig.~\ref{EnDensityQM}.

\subsection{Spinball plus a quantum dyon}
\label{QDS}

\begin{figure}[b]
	\begin{minipage}[t]{.45\linewidth}
		\begin{center}
			\includegraphics[width=1\linewidth]{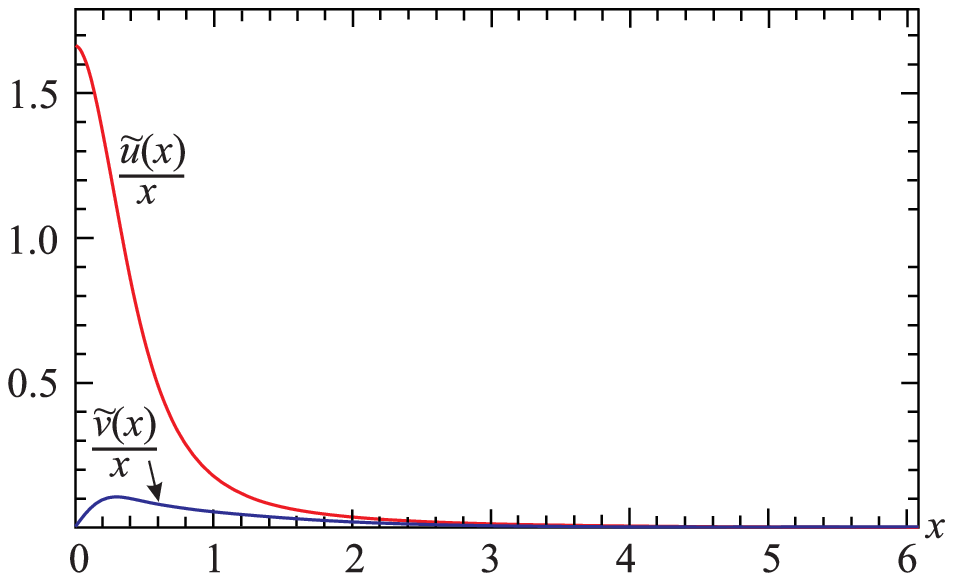}
		\end{center}
\vspace{-0.5cm}		
\caption{The graphs of the functions $\tilde v(x)$ and $\tilde u(x)$ of the solution describing the
 spinball-plus-quantum-dyon system at
		$\tilde \lambda = 0.1$,
		$\tilde \Lambda = 1/9$, $\tilde m_q = 1$, $m = 3$, $\tilde g=1$,
		$\tilde \xi_0 = 0.5$, $f_2 = - 20$, $\tilde \chi_0 = 0$, and $\tilde E = 0.8$.
The corresponding eigenvalues are $\tilde M \approx 1.275$, $\tilde \mu \approx 2.6725351952$, and
		$\tilde u_1 \approx 1.6653$.
		}
		\label{fphi_dyon}
	\end{minipage}\hfill
	\begin{minipage}[t]{.45\linewidth}
		\begin{center}
			\includegraphics[width=1\linewidth]{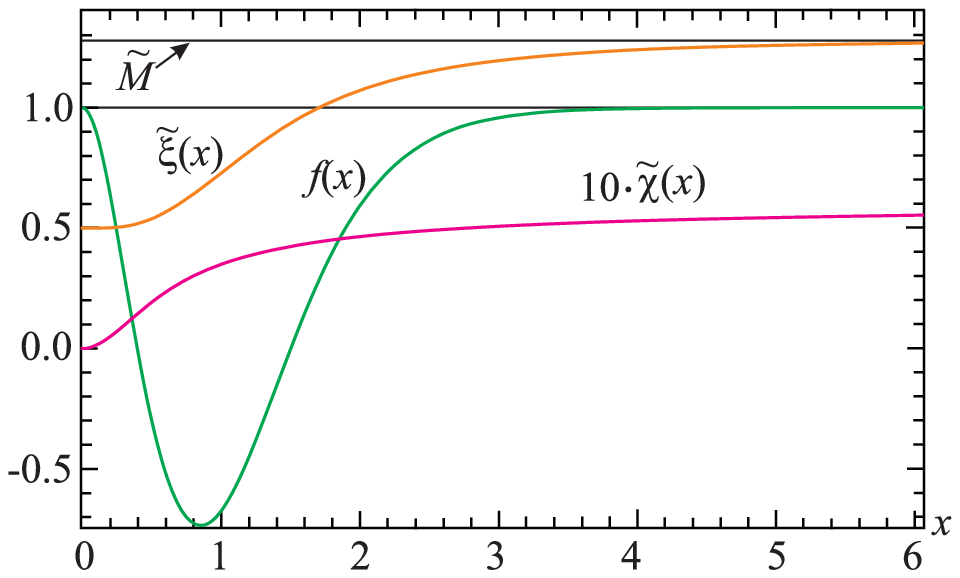}
		\end{center}
\vspace{-0.5cm}		
		\caption{	
			The profiles of the functions $\tilde \xi(x),  f(x)$, and $\tilde \chi(x)$ for the values of the parameters given in caption to Fig.~\ref{fphi_dyon}.
		}
		\label{fvchi_dyon}
	\end{minipage}
\end{figure}

\begin{figure}[t]
		\begin{center}
			\includegraphics[width=.45\linewidth]{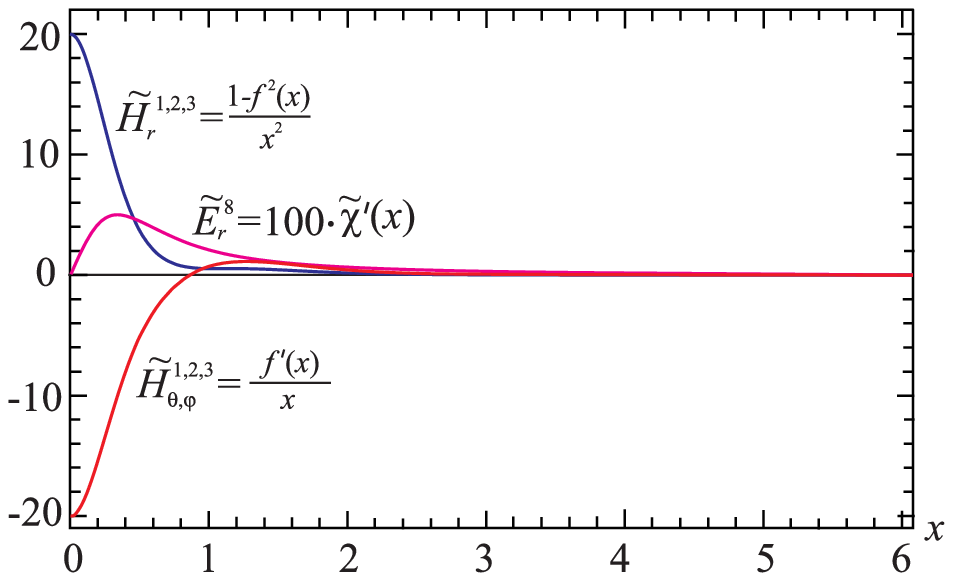}
		\end{center}
\vspace{-0.5cm}		
		\caption{	
			The profiles of the color magnetic, $\tilde H^{1,2,3}_r$,
			$\tilde H^{1,2,3}_{\theta, \varphi}$, and electric, $\tilde E^8_r$, fields
			for the spinball-plus-quantum-dyon configuration.
		}
		\label{fields_dyon}
\end{figure}

Here, we consider the complete set of equations  \eqref{2-40}-\eqref{2-70} describing a quantum system with color electric and magnetic fields
(a quantum dyon) plus quarks. Similarly to the previous sections, we seek a solution of these equations as $x \rightarrow 0$  in the following form:
\begin{eqnarray}
	f &=& 1 + \frac{f_2}{2} x^2 + \ldots ,
\label{3-d-60}\\
	\tilde \chi &=&\tilde \chi_0+ \tilde \chi_2 \frac{x^2}{2} + \ldots ,
\label{3-d-55}\\
	\tilde \xi &=& \tilde \xi_0 + \frac{\tilde \xi_2}{2} x^2 + \ldots ,
\label{3-d-70}\\
	\tilde u &=& \tilde u_1 x + \frac{\tilde u_3}{3!} x^3 + \ldots ,
\label{3-d-80}\\
	\tilde v &=& \frac{\tilde v_2}{2} x^2 + \frac{\tilde v_4}{4!} x^4 + \ldots , \quad \text{where} \quad
	\tilde v_2 = \frac{2}{3} \tilde u_1 \left(
		\tilde E-\tilde m_q  + m^2 \tilde\Lambda \tilde\xi_0  \tilde u_1^2+\frac{\tilde \chi_0}{2 \sqrt{3}}
	\right) .
\label{3-d-90}
\end{eqnarray}
As before, Eqs.~\eqref{2-40}-\eqref{2-70} are solved numerically as a nonlinear problem for the eigenvalues $\tilde \mu, \tilde M, \tilde E$ and the eigenfunctions $f(x), \tilde \chi, \tilde \xi(x), \tilde v(x), \tilde u(x)$.
Examples of the corresponding solutions are given in Figs.~\ref{fphi_dyon} and \ref{fvchi_dyon}. Their asymptotic behavior
as $x\to \infty$ is as follows:
\begin{eqnarray}
	f(x) &\approx& 1 - f_\infty e^{- x \sqrt{m^2 \tilde{M}^2 - \tilde\mu^2}} , \quad
\tilde \chi(x) \approx \tilde \chi_\infty - \frac{Q}{x} , \quad
\tilde\xi (x) \approx \tilde M -
	\tilde\xi_\infty \frac{e^{- x \sqrt{2 \tilde\lambda \tilde{M}^2}}}{x},
\label{3-d-100}\\
\tilde u &\approx& \tilde u_\infty e^{- x \sqrt{
			\tilde m_q^2 - \tilde E^2
	}} ,\quad
\tilde v \approx \tilde v_\infty e^{- x \sqrt{
	\tilde m_q^2 - \tilde E^2
	}}.
\label{3-d-120}
\end{eqnarray}
Here $f_\infty, \tilde \chi_\infty, Q, \tilde \xi_\infty, \tilde v_\infty$, and $\tilde u_\infty$ are some constants. The corresponding distributions of different components of the color magnetic and electric fields are shown in Fig.~\ref{fields_dyon}.

The energy density of the quantum dyon embedded in the condensate is given by the expression \eqref{2-120}, and its graph practically coincides with that of the energy density of the quantum  monopole shown in Fig.~\ref{EnDensityQM}
(for the same values of the free parameters).

It must be emphasized here that there is a crucial difference between  asymptotic behaviors of the color magnetic and electric fields
(see Eq.~\eqref{3-d-100} and Fig.~\ref{fields_dyon}): the magnetic field decreases according to an exponential law and the electric field -- according to a power law (the Coulomb law).

\section{Mass gap in the nonperturbative quantization \`a la Heisenberg}
\label{MassGap}

In this section we consider energy spectra of two systems (spinball and  spinball-plus-quantum-monopole ones) and show the existence of mass gaps in them.

\subsection{Mass gap for the  spinball}
\label{MassGapSB}

\begin{figure}[t]
	\begin{minipage}[t]{.45\linewidth}
		\begin{center}
			\includegraphics[width=1\linewidth]{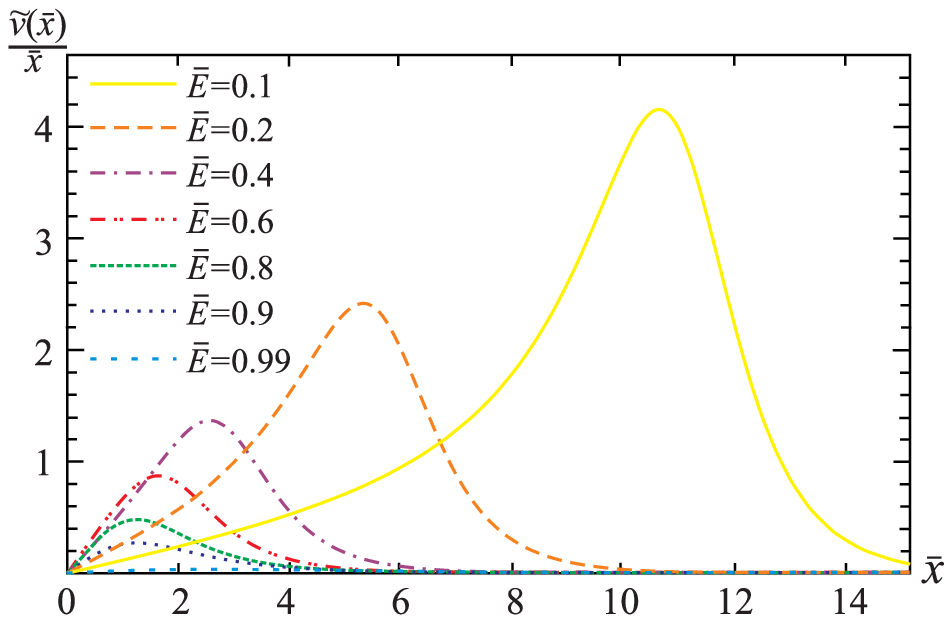}
		\end{center}
\vspace{-0.5cm}		
\caption{Family of the spinball solutions $\tilde v(\bar x)$ for different values of the parameter $\bar E$.
		}
		\label{v_x_family}
	\end{minipage}\hfill
	\begin{minipage}[t]{.45\linewidth}
		\begin{center}
			\includegraphics[width=1\linewidth]{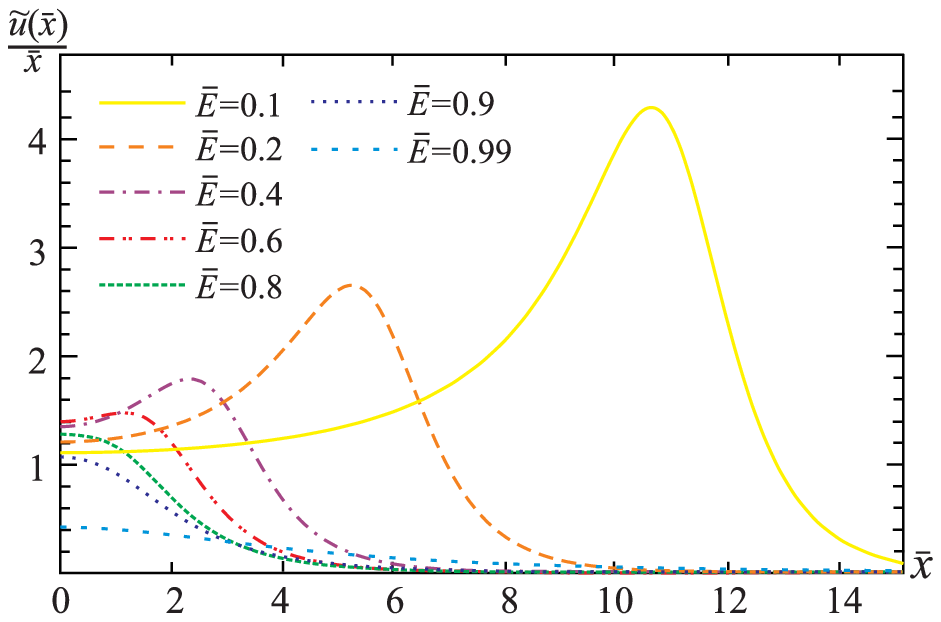}
		\end{center}
\vspace{-0.5cm}
		\caption{	
			Family of the spinball solutions $\tilde u(\bar x)$ for different values of the parameter $\bar E$.
		}
		\label{u_x_family}
	\end{minipage}
	\begin{minipage}[t]{.45\linewidth}
		\begin{center}
			\includegraphics[width=1\linewidth]{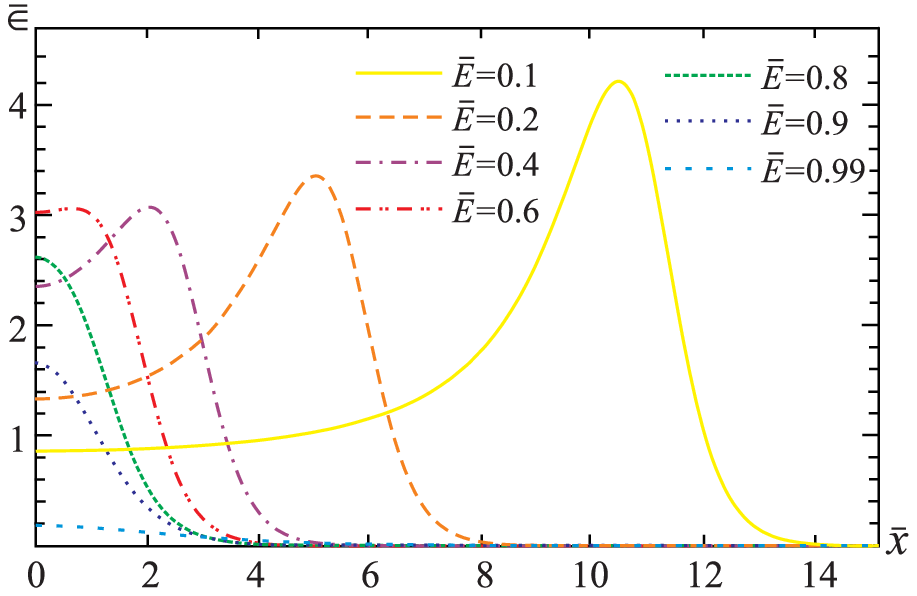}
		\end{center}
\vspace{-0.5cm}		
\caption{	
			Family of  the spinball energy densities $\bar \epsilon(\bar x)$ for different values of the parameter $\bar E$.
		}
		\label{EnDenSB_QM}
	\end{minipage}\hfill
	\begin{minipage}[t]{.45\linewidth}
		\begin{center}
			\includegraphics[width=1\linewidth]{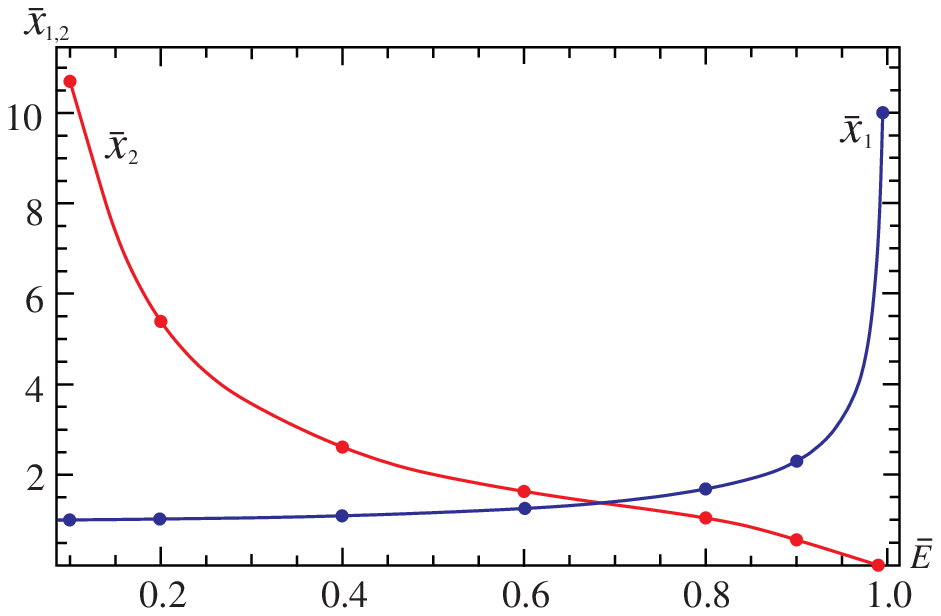}
		\end{center}
\vspace{-0.5cm}	
	\caption{The dependence of the characteristic sizes of the spinball for the left- and right-hand branches of the energy spectrum shown in Fig.~\ref{massgapSB}.
		}
		\label{x12}
	\end{minipage}
\end{figure}

Let us calculate an energy spectrum of the spinball described by  Eqs.~\eqref{2-a-20} and \eqref{2-a-30}. In these equations, there is only one parameter, $\bar E$, giving the energy. Two remaining parameters ($\Lambda$ and $\tilde g$) are the constants of the theory, and they cannot change.
Solving these equations for different values of the parameter $\bar E$, we get the data given in Table~\ref{mgsp}. Figs.~\ref{v_x_family} and \ref{u_x_family} show the families of the corresponding solutions for the functions $\tilde v(x)$ and $\tilde u(x)$.

\begin{table}[h]
\scalebox{1.}{
	\begin{tabular}{|c|c|c|c|c|c|c|c|c|c|}
	\hline
	\rule[-1ex]{0pt}{2.5ex}
	$\bar E$&0.1&0.2&0.4&0.6&0.8&0.9&0.99&0.999&0.9999\\
	\hline
	\rule[-1ex]{0pt}{2.5ex}
	$\tilde u_1$&1.103640998&1.20242169&1.34371184&1.389621&1.2745556&1.06477&
		0.419164&0.1366701&0.0433583\\
	\hline
	\rule[-1ex]{0pt}{2.5ex}
	$\bar W_t$&18585&2985.04&470.563&152.677&66.4478&49.285&
	74.2536&213.603&668.854\\
	\hline
    \end{tabular}
	}
\caption{Eigenvalues $\tilde u_1$ and the total energy $\bar W_t$ from \eqref{4-a-20} as functions of the parameter $\bar E$. }
\label{mgsp}
\end{table}

\begin{figure}[t]
	\begin{minipage}[t]{.45\linewidth}
		\begin{center}
			\includegraphics[width=1\linewidth]{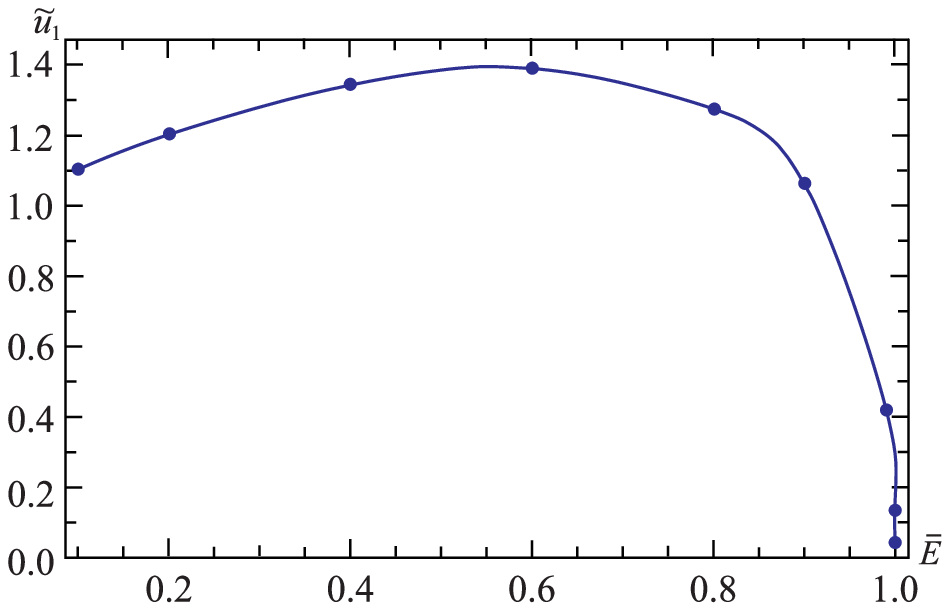}
		\end{center}
\vspace{-0.5cm}		
\caption{The dependence of the eigenvalue $\tilde u_1$ on $\bar E$ for the spinball.
		}
		\label{u1TdE}
	\end{minipage}\hfill
	\begin{minipage}[t]{.45\linewidth}
		\begin{center}
			\includegraphics[width=1\linewidth]{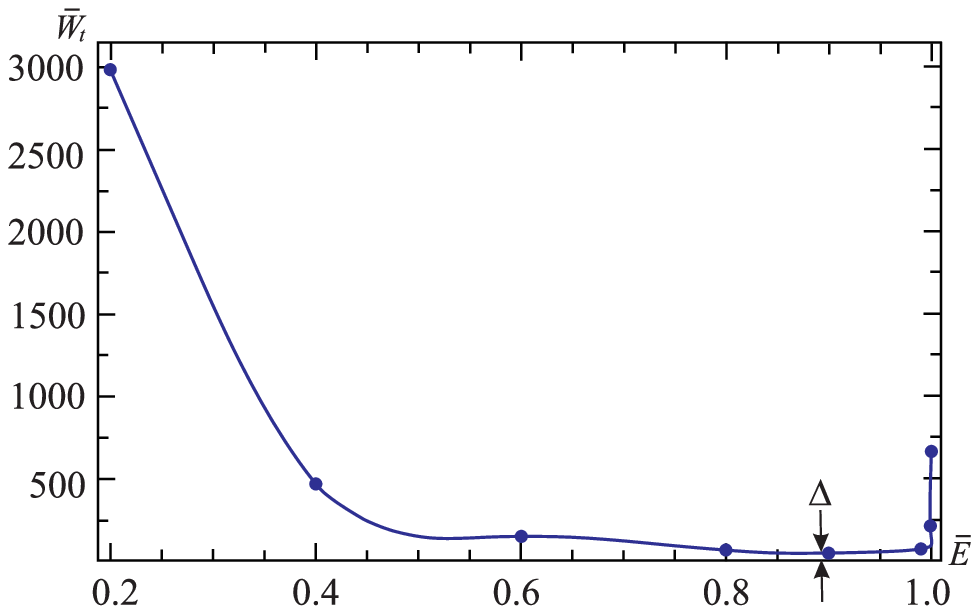}
		\end{center}
\vspace{-0.5cm}		
\caption{The dependence of the spinball total energy	$\bar W_t$ on $\bar E$ and the location of the  mass gap	$\Delta$.
		}
		\label{massgapSB}
	\end{minipage}
\end{figure}

The dimensionless energy density for the spinball under consideration has the following form:
\begin{equation}
	\tilde \epsilon(\bar x) = \tilde m_q^3 \left[
	\bar E \frac{\tilde u^2 + \tilde v^2}{\bar x^2} +
			\frac{\bar \Lambda}{2 }
			\frac{\left(\tilde u^2 - \tilde v^2 \right)^2}{\bar x^4}
	\right] =\tilde m_q^3\bar \epsilon(\bar x).
\label{4-a-10}
\end{equation}
The corresponding profiles of  $\bar \epsilon(\bar x)$ are shown in  Fig.~\ref{EnDenSB_QM}. The numerical analysis indicates that as
$\bar E \rightarrow 1$ the maximum of the energy density is located at $\bar x = 0$ and goes to 0 (see Fig.~\ref{EnDenSB_QM}), and the linear size of the configuration becomes infinitely large. In this case, according to the numerical calculations, the total energy of the  spinball \eqref{4-a-20} tends to infinity. In turn, as $\bar E \rightarrow 0$, it seems that the maximum of the energy density goes to infinity, and the location of the maximum along the radius shifts to infinity as well. This is also approved by the results of numerical calculations of the total energy  \eqref{4-a-20} given in Fig.~\ref{massgapSB}.

Using Eq.~\eqref{4-a-10}, the total dimensionless energy $\tilde W_t$ of the system under consideration can be calculated as
\begin{equation}
	\tilde W_t \equiv W_t/\left(\hbar c/r_0\right)= 4 \pi \tilde m_q^3
	\int\limits_0^\infty \bar \epsilon \bar x^2  d\bar x =
	\tilde m_q^3 \bar W_t.
\label{4-a-20}
\end{equation}

The dependencies $\tilde u_1(\bar E)$ and $\bar W_t(\bar E)$ are given in Figs.~\ref{u1TdE} and \ref{massgapSB}. It is seen from Fig.~\ref{massgapSB} that at some value of the parameter $\bar E$ there exists a minimum value of the total energy  $\bar W_t$. This corresponds to the presence of the mass gap $\Delta$ for the spinball in question. Apparently, a mass gap was first found for the nonlinear Dirac equation \eqref{1-140} (without the condensate $\phi$) in Refs.~\cite{Finkelstein:1951zz,Finkelstein:1956} (in the terminology of the authors, that was ``the lightest stable particle'').

\subsection{Mass gap for the spinball-plus-quantum-monopole system}
\label{mgsbqm}

Consider now more realistic case of the spinball-plus-quantum-monopole system. This is a ball filled with spinor and color magnetic fields embedded in the condensate. Its structure is described by Eqs.~\eqref{3-c-10}-\eqref{3-c-40}, using which, in Sec.~\ref{QMS} we have constructed the corresponding regular solutions. Here, we find the energy spectrum of such system and show the presence of the mass gap.

Using the expression for the dimensionless energy density $ \tilde \epsilon $ from Eq.~\eqref{3-c-130}, the dimensionless total energy of the system in question is calculated as
 \begin{equation}
	\tilde W_t \equiv \frac{W_t}{ \hbar c/r_0  } = 4 \pi
	\int\limits_0^\infty x^2 \tilde \epsilon d x.
\label{4-b-10}
\end{equation}
Solving the set of equations \eqref{3-c-10}-\eqref{3-c-40} numerically, we have computed this energy for different values of $f_2$ and $\tilde E$ (see in Appendix~\ref{app1}).

\begin{figure}[t]
	\begin{minipage}[t]{.45\linewidth}
		\begin{center}
			\includegraphics[width=1\linewidth]{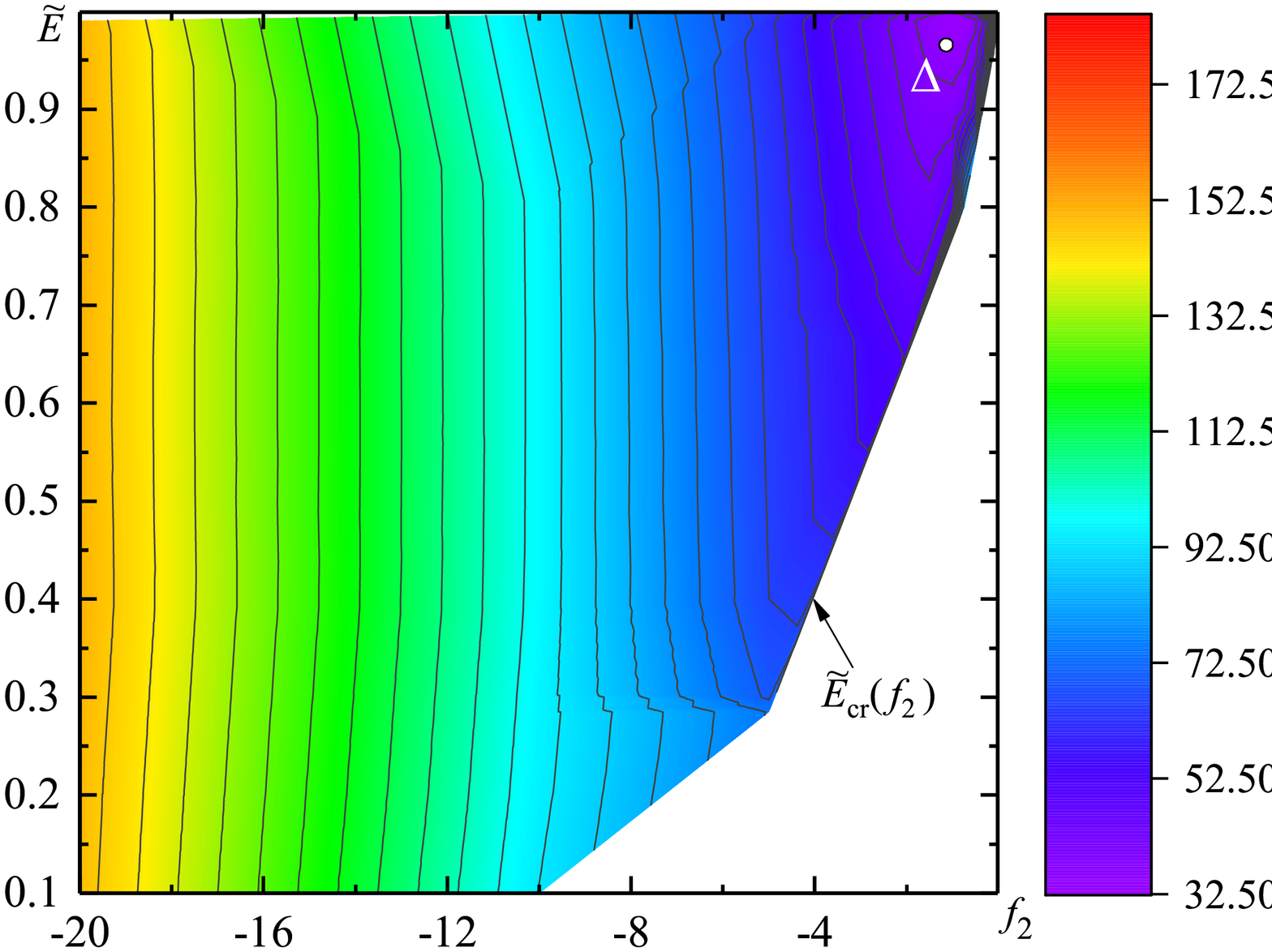}
		\end{center}
\vspace{-0.5cm}		
\caption{Contour plot of the total energy $\tilde W_t(\tilde E, f_2)$ from Eq.~\eqref{4-b-10}
(the data are given in Appendix~\ref{app1}).
$\Delta$ shows the location of the mass gap in the $\left\lbrace f_2, \tilde E \right\rbrace$ plane.}
		\label{DensityPlotMG1}
	\end{minipage}\hfill
	\begin{minipage}[t]{.45\linewidth}
		\begin{center}
			\includegraphics[width=1\linewidth]{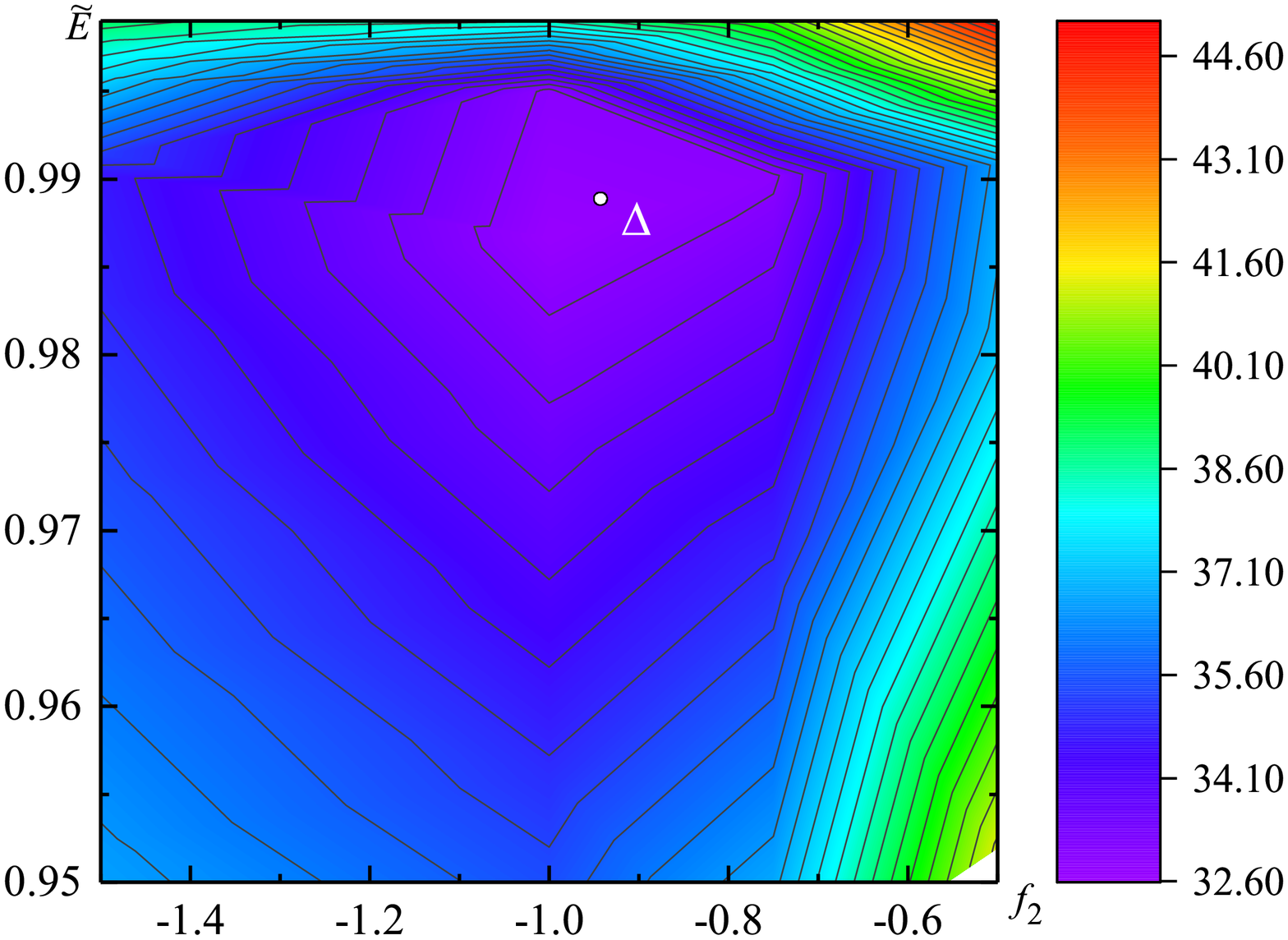}
		\end{center}
\vspace{-0.5cm}
		\caption{Contour plot of the total energy $\tilde W_t(\tilde E, f_2)$ from Eq.~\eqref{4-b-10}
in the vicinity of the point where the mass gap $\Delta$ is located. }
		\label{DensityPlotMG2}
	\end{minipage}
	\begin{minipage}[t]{.45\linewidth}
		\begin{center}
			\includegraphics[width=1\linewidth]{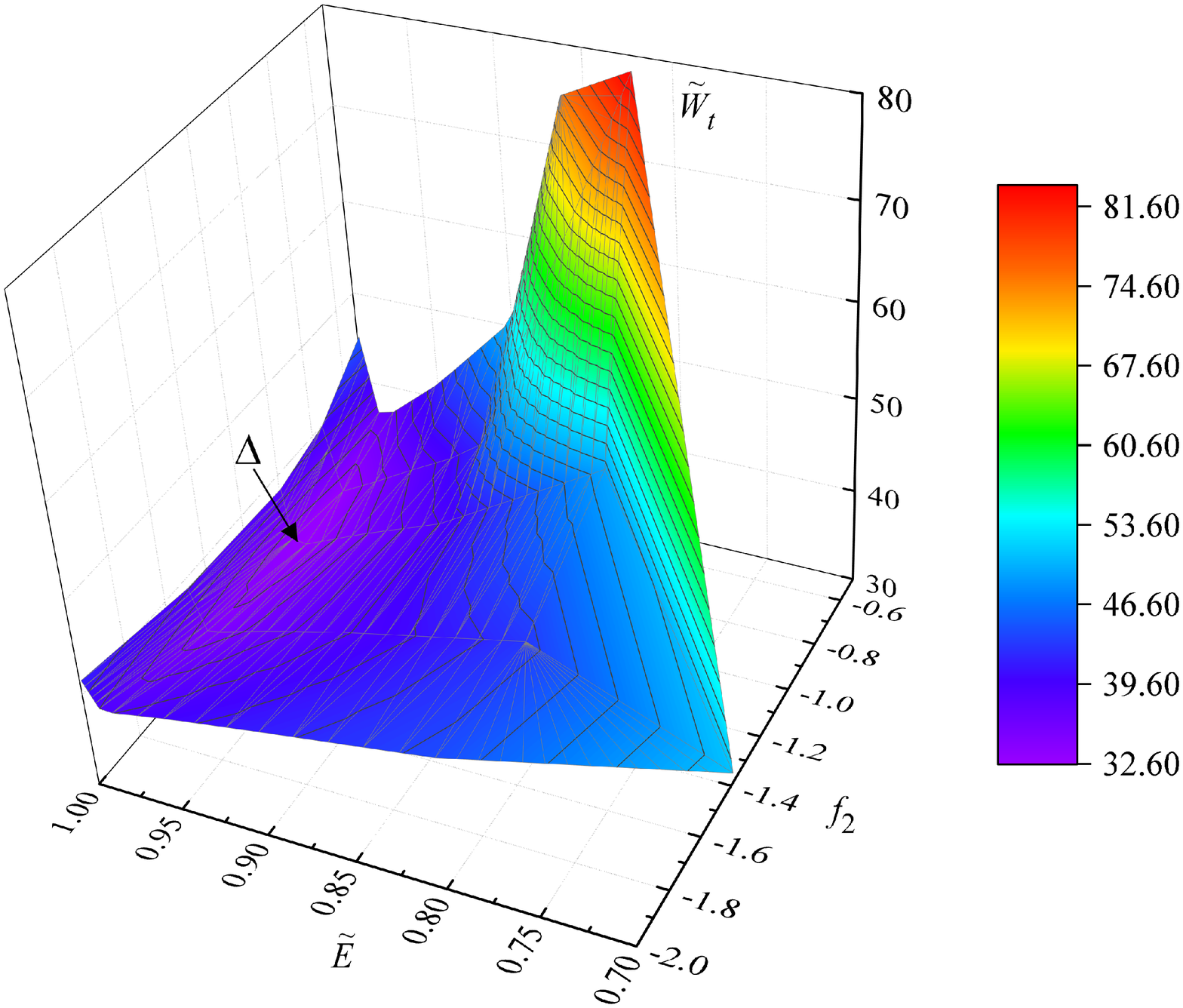}
		\end{center}
\vspace{-0.5cm}
		\caption{Three-dimensional plot for the dependence of the total energy \eqref{4-b-10} on the parameters
		$ f_2$ and $\tilde E $ in the vicinity of the mass gap $\Delta$.}
		\label{3Dplot}
	\end{minipage}\hfill
	\begin{minipage}[t]{.45\linewidth}
		\begin{center}
			\includegraphics[width=1\linewidth]{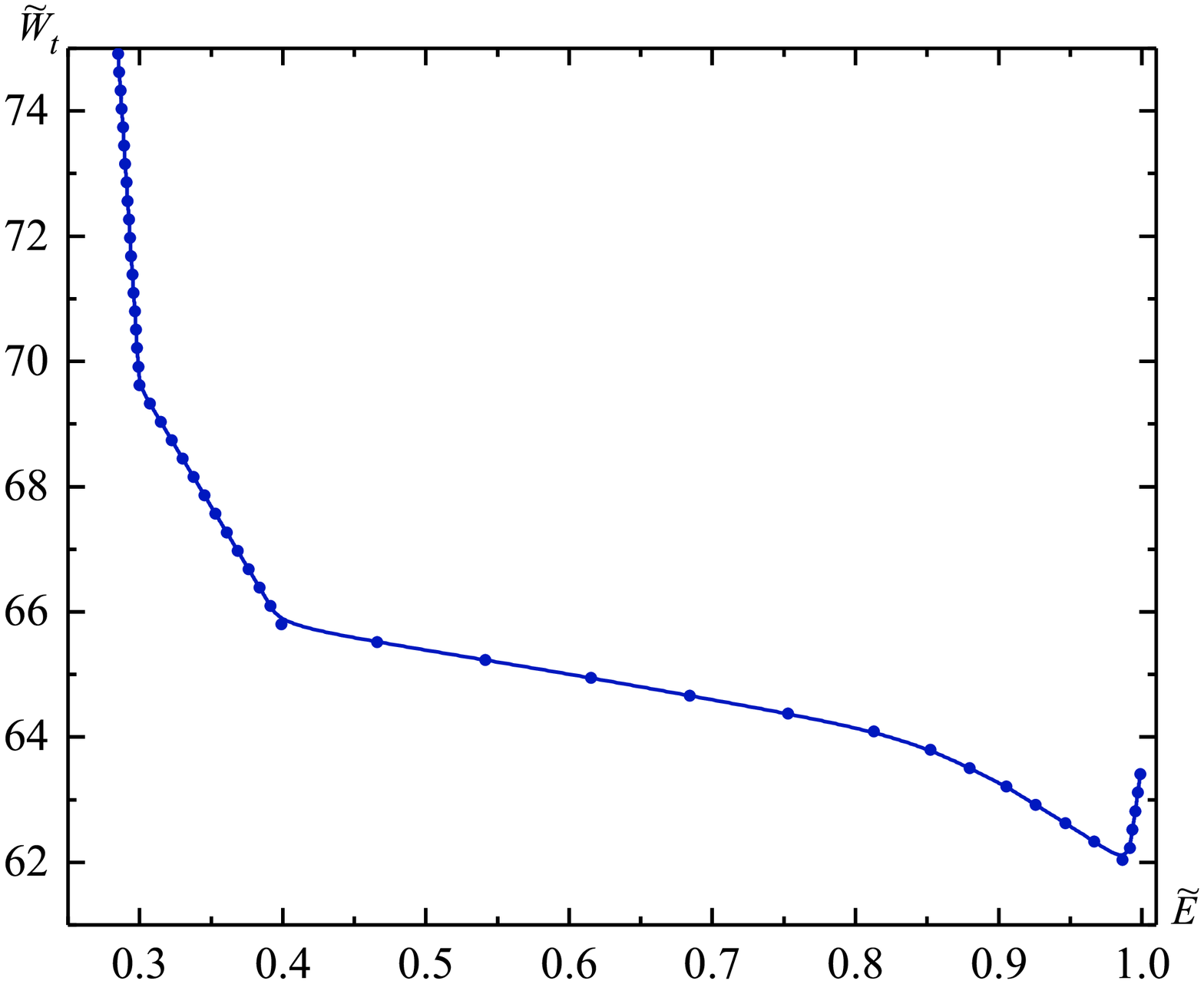}
		\end{center}
\vspace{-0.5cm}		
\caption{The dependence of the total energy \eqref{4-b-10} on $\tilde E$ at fixed $f_2 = -5$.}
		\label{Wt_E}
	\end{minipage}
\end{figure}

Three-dimensional and contour plots for the energy \eqref{4-b-10} are given in Figs.~\ref{DensityPlotMG1}-\ref{3Dplot}. Figure~\ref{DensityPlotMG2} shows the region of Fig.~\ref{DensityPlotMG1} in the vicinity of the energy minimum, i.e., the region where the mass gap is resided. One can see from this figure that there are closed lines characterising the presence of the extremum of the energy. In the case under consideration, this is the minimum corresponding to the mass gap. The approximate value of the dimensionless mass gap for the magnitudes of the parameters used here
($\tilde \Lambda = 1/9,\tilde  m_q = 1, \tilde g = 1$, and $m = 3$) is
\begin{equation}
	\frac{\Delta}{\hbar c/r_0} = \tilde \Delta \approx 32 \quad \text{ at } \quad
	f_2 \approx -0.94, \; \tilde E \approx 0.99 ,
\label{4-b-20}
\end{equation}
where $r_0$ is of the order of  $\Lambda_{\text{QCD}}$, as it is explained in Sec.~\ref{LambdaQCD}.

Notice here the following important features of the system under consideration:
\begin{itemize}
  \item The numerical analysis indicates that as $\tilde E \rightarrow 1$ the total energy $\tilde W_t \rightarrow + \infty$.
  \item Regular solutions to Eqs.~\eqref{3-c-10}-\eqref{3-c-40} exist not for all pairs $\left\lbrace f_2, \tilde E \right\rbrace $.
  It is seen from Fig.~\ref{DensityPlotMG1} that for some fixed  $f_2$ there exists a critical value $\tilde E_{\text{cr}}$
  at which the solution still exists but no solutions are found for $\tilde E < \tilde E_{\text{cr}}$.
  \item The determination of  the asymptotic behavior of $\tilde W_{t}$ as $\tilde E \rightarrow \tilde E_{\text{cr}} $
  offers formidable computing difficulties. According to the numerical studies, we can assert rather confidently
  that in the limit $\tilde E \rightarrow \tilde E_{\text{cr}} $  the value of the total energy
  $\tilde W_t$ (at fixed $f_2 = \text{const.}$) does not, at least, decrease, i.e., its boundary value goes either to
   $+\infty$ or to some constant value depending on $f_2$ only. As an example, Fig.~\ref{Wt_E}
   shows the profile of the total energy for $f_2 = -5$. It confirms that asymptotically, as $\tilde E \rightarrow 1$, the total energy diverges.
   Similar behavior of the total energy is also possible as $f_2 \rightarrow f_{2, \text{cr}}$.
\end{itemize}
Hence, one can see from the above that regular solutions to the set of equations \eqref{3-c-10}-\eqref{3-c-40} in the
$\left\lbrace f_2, \tilde E \right\rbrace $ plane do exist in the region restricted by the line $\tilde E \to 1$ and the critical curve $\tilde E_{\text{cr}} = \tilde E_{\text{cr}}(f_2)$ (shown in Fig.~\ref{DensityPlotMG1}).

\section{Deconfinement}
\label{deconf}

In this section we give a qualitative discussion of  the deconfinement mechanism when one takes into account the nonperturbative effects.

\subsection{Spinball}

Based on the results concerning the spinball energy spectrum obtained in Sec.~\ref{MassGapSB}, it is possible to propose the following deconfinement mechanism. As is seen from Fig.~\ref{massgapSB}, the energy of the spinball can be arbitrarily large. This results in the fact that with rising temperature the average energy of the spinball will increase. The typical size of the spinball for the right-hand branch of the energy spectrum, according to the asymptotic expressions~\eqref{2-a-80}, is defined as follows:
\begin{equation}
	\bar x_1 \sim \frac{1}{\sqrt{ 1 - \bar E^2}} .
\label{5-10}
\end{equation}
It is seen from Fig.~\ref{massgapSB} that on the right-hand branch of the energy spectrum $\bar W_t \xrightarrow{\bar E \rightarrow 1} \infty$. Thus, with rising temperature, the typical size of the spinball on the right-hand branch will tend to infinity.

In turn, the typical size of the  spinball $\bar x_2$ for the left-hand branch of the energy spectrum is defined as the location of maxima of the energy density given in Fig.~\ref{EnDenSB_QM}. The corresponding behavior of $\bar x_{1,2}$ as functions of $\bar E$ is shown in Fig.~\ref{x12}.

Given the spinball density  $n_s$, the typical distance between spinballs is defined as follows:
\begin{equation}
	l_s \sim \left( n_s \right)^{-1/3} .
\label{5-20}
\end{equation}
At $\bar x_{1,2} \approx l_s$ the spinball solution of Sec.~\ref{spinball}
becomes incorrect. In this case quarks would not already be confined in a restricted region, and one has to consider another solution for the quarks.

Thus, one may suppose that quark deconfinement takes place at
\begin{equation}
	\bar x_{1,2} \sim l_s .
\label{5-30}
\end{equation}
Here, the typical size of the spinball depends on the temperature  $T$.
Notice that since in general case $\bar x_1 \neq \bar x_2$, deconfinement for the left- and right-hand sides of the energy spectrum will occur at different temperatures. This means that the transition from the confinement phase to the full deconfinement phase may happen in two stages. On the first stage, deconfinement occurs either for the left-hand branch or for the right-hand branch, and then deconfinement occurs for another branch of the energy spectrum.

\subsection{Spinball plus quantum monopole}

The same process will also take place for the spinball-plus-quantum-monopole configuration. According to the results obtained in  Sec.~\ref{mgsbqm} for
the dependence of the total energy of the spinball-plus-quantum-monopole system on the parameters $f_2$ and $\tilde E$, the energy of such an object may also be infinite (see Figs.~\ref{DensityPlotMG1}-\ref{Wt_E}). In this case the typical size of the configuration is determined by the asymptotic formulae
\eqref{3-c-90} and \eqref{3-c-110}. One might expect that a qualitative behavior of characteristic sizes of such configurations would be similar to that of the pure spinball considered above. But this issue requires additional study that we intend to address in future work.

\section{Dimensional transmutation, quantum correlators in a spacelike direction, and $\Lambda_{\text{QCD}}$}
\label{LambdaQCD}

Let us now discuss how the constant $\Lambda_{\text{QCD}}$ appears in QCD from using the nonperturbative approach \`a la Heisenberg. As is known, the
Feynman propagator (2-point Green's function), for example, for a real scalar field, is defined as follows:
\begin{equation}
	\Delta_F \left( x^\mu - y^\mu \right) = \left\langle
		T \hat \phi(x^\mu) \hat \phi(y^\mu)
	\right\rangle,
\label{4-10}
\end{equation}
and it is equal to zero for spacelike directions. Within the three-equation approximation used here, we deal, for instance, with the following 2-point Green functions:
\begin{eqnarray}
	G^{mn \mu \nu}(y,x) &=& \left\langle
		\hat A^{m \mu}(y) \hat A^{n \nu}(x)
	\right\rangle \approx C^{mn \mu \nu} \phi(y) \phi(x) ,
\label{4-20}\\
	G^{ab \mu \nu}(y,x) &=& \left\langle
		\widehat{\delta A}^{a \mu}(y) \widehat{\delta A}^{b \nu}(x)
	\right\rangle \approx C^{ab \mu \nu}.
\label{4-30}
\end{eqnarray}
Here $a, b = 1,2,3, 8$ and $m,n = 4,5,6,7$; $C^{mn \mu \nu}$ and $C^{ab \mu \nu}$ are some numerical coefficients. These Green functions, unlike the Feynman propagator \eqref{4-10} for a free field, are not equal to zero for spacelike directions. In Eqs.~\eqref{1-10} and \eqref{1-20},
we have the following quantities:
\begin{eqnarray}
	\left( m^2 \right)^{ab \mu\nu} &=& - g^2 \left[
		f^{abc} f^{cpq} G^{pq \mu\nu} -
		f^{amn} f^{bnp} \left(
			\eta^{\mu \nu} G^{mp \phantom{\alpha} \alpha}_{\phantom{mn} \alpha} -
			G^{mp \nu \mu}
		\right)
	\right] ,
\label{4-40}\\
\left( \mu^2 \right)^{ab \mu \nu} &=& - g^2 \left(
		f^{abc} f^{cde} G^{de \mu \nu} +
		\eta^{\mu \nu} f^{adc} f^{cbe} G^{de \phantom{\alpha} \alpha}_{\phantom{de} \alpha} +
		f^{aec} f^{cdb} G^{ed \nu \mu}
	\right) ,
\label{4-50}\\
	\left( m^2_\phi \right)^{ab \mu \nu} &=&
	g^2 f^{amn} f^{bnp} \frac{
		G^{mp \mu \nu} - \eta^{\mu \nu} G^{mp \alpha}_{\phantom{mp \alpha} \alpha}
	}
	{G^{mm \alpha}_{\phantom{mm\alpha} \alpha}}	.
\label{4-60}
\end{eqnarray}
For these Green functions, we use the  approximations
\begin{eqnarray}
	G^{mn \mu \nu}(y,x) &\approx&
	\Delta^{mn} \mathcal A^\mu \mathcal A^\nu \phi(y) \phi(x) ,
\label{4-70}\\
	G^{ab \mu \nu}(y,x) &\approx&
	\Delta^{ab}	\mathcal B^\mu \mathcal B^\nu ,
\label{4-80}
\end{eqnarray}
where $\Delta^{ab} (a,b = 1,2,3,8), \Delta^{mn} (m,n = 4,5,6,7)$ are constants and $\mathcal A^\mu, \mathcal B^\mu$ are some vectors (for details see Ref.~\cite{Dzhunushaliev:2016svj}). These approximations, along with the ansatz \eqref{2-10}-\eqref{2-30}, permit obtaining Eqs.~\eqref{2-40}-\eqref{2-70} with the constants $\mu = \tilde \mu / r_0$ and $M = \tilde M / (g r_0)$.

The asymptotic behavior of the 2-point Green function \eqref{4-20} is determined by the asymptotic behavior \eqref{2-b-50} and \eqref{3-c-90} of the function $\phi(x)$ describing the condensate of the coset field $A^m_\mu \in \text{SU(3)}/ \left( \text{SU(2)} \times \text{U(1)} \right)$. This means that the exponential drop of this Green function is determined by the parameter $r_0$. Thus we have shown that the 2-point Green function \eqref{4-20} refers to some dimensional quantity $r_0$ which has the dimensions of length. This enables one to relate this constant to the following constant known from QCD: $r_0 = \Lambda_{\text{QCD}} \approx 10^{-15}~\text{m}$.

This allows us the possibility of drawing a very important conclusion:
\emph{in quantizing strongly nonlinear fields, Green's functions for spacelike directions are not equal to zero and they are determined by some characteristic dimensional quantity.} In QCD, $\Lambda_{\text{QCD}}$ is such a quantity.
Apparently, such quantities are not new fundamental constants but depend on the specific form of the Lagrangian of a theory in question.

\section{Physical applications and conclusions}
\label{concl}

The physical systems considered in the present paper represent the quantum condensate, which is described by the function $\phi$ and contains either one [when $\tilde \epsilon_\infty = 0$ from Eq.~\eqref{2-120}] or some quantity  (when $\tilde \epsilon_\infty \neq 0$) of spinballs, quantum monopoles, spinball-plus-quantum-monopole configurations, and spinball-plus-quantum-dyon systems. In the first case, we assume that one can say about an approximate
glueball model (a quantum monopole or a spinball-plus-quantum-monopole system). In the second case, it is assumed that there is an approximate description of the QCD vacuum or a quark-gluon plasma filled with quantum fluctuations, each of which is either a spinball, or a quantum monopole, or a spinball-plus-quantum-monopole system, or a spinball-plus-quantum-dyon configuration. When two quantum monopoles with opposite magnetic charges are present, they will create a monopole-antimonopole pair, similar to a Cooper pair in a superconductor. Particles in such a pair will be connected by a flux tube filled with color magnetic and electric fields. Note that within the framework of the two-equation approximation in the nonperturbative quantization \`a la Heisenberg one can obtain an infinitely long flux tube~\cite{Dzhunushaliev:2017rin}. The obtaining a finite-length flux tube runs into great difficulty, since in this case one has to solve a nonlinear eigenvalue problem for a set of partial differential equations.

Thus, the following results have been obtained:
\begin{itemize}
	\item The energy spectra for the spinball and for the 				   spinball-plus-quantum-monopole system.
	It was shown that they possess the mass gaps.
	\item Within the framework of the three-equation approximation,
	solutions describing virtual quarks and gauge fields in a bag have been found.
	\item It was shown that the bags are created due to the Meissner effect, when 	the  coset condensate expels the gauge fields.
 	\item For the quantum-monopole/dyon systems, it was shown that color magnetic and electric fields decrease asymptotically according to
 	exponential and power laws, respectively.
 	\item A comparison of some solutions obtained within the framework of the given approximation with lattice and phenomenological approximate calculations has been carried out.
  	\item The qualitative deconfinement model has been suggested, according to which at some temperature a characteristic size of a
 	spinball or of a spinball-plus-quantum-monopole configuration becomes comparable to a characteristic distance between them.
 	\item It was shown that the QCD constant $\Lambda_{\text{QCD}}$ appears in the nonperturbative quantization
 	\`a la Heisenberg as some constant controlling the correlation length of quantum fields in a spacelike direction.
 	\item The nonlinear Dirac equation has been used as an approximate description of an infinite set of equations for all Green functions of the spinor equation.
 \end{itemize}
It is interesting to note that in the 1950's the mass gap has in fact been found in Refs.~\cite{Finkelstein:1951zz,Finkelstein:1956} in solving the nonlinear Dirac equation. However, the authors did not use such term, but
said of ``the lightest stable particle''. Those papers were devoted to study of the nonlinear Dirac equation, and  W.~Heisenberg offered to use it as a fundamental equation in describing the properties of an electron.
In our case, we employ that equation in describing the dispersion
$\left\langle \hat{\bar \psi} \hat \psi \right\rangle$ of the quark field and the correlation
$\left\langle \hat{\bar \psi} \hat A^m_\mu \hat \psi \right\rangle$ between the spinor and gauge fields having a zero quantum average:
$\left\langle \hat \psi \right\rangle = 0$ and
$\left\langle \hat A^m_\mu \right\rangle = 0$.
To the best of our knowledge, the mass gap was first obtained in Refs.~\cite{Finkelstein:1951zz,Finkelstein:1956}.

Also, note that the parameters  $\left( m^2 \right)^{ab \mu \nu}$,
$\left( \mu^2 \right)^{ab \mu \nu}$, and $\left( m^2_\phi \right)^{ab \mu \nu}$ are similar to the closure parameters known in turbulence modeling as the parameters appearing when one cuts off an infinite set of equations coming from the Navier-Stokes equation in stochastic averaging~\cite{Wilcox}.

In conclusion, we emphasise that the calculations performed here are only valid for a stationary case.

\section*{Acknowledgements}

This work was supported by Grant No.~BR05236730 in Fundamental Research in Natural Sciences by the Ministry of Education and Science of the Republic of Kazakhstan.

\appendix

\section{The values of the total energy $\tilde W_t$ and of the eigenvalues  $\mu, M, u_1$ at different $f_2, \tilde E$ for the
spinball-plus-quantum-monopole system}
\label{app1}


\vspace{.2cm}

\begin{longtable}{|c|c|c|c|c|c|c|c|c|c|c|}
\hline
\multicolumn{11}{|c|}{$f_2 = - 0.02$} \\
\hline
 $\tilde E$	& 0.999 	& 0.99 		&  &  &  &  &  & & & \\
\hline
 $\tilde W_t$			& 184.244 & 135.626 &  &  &  &  &  & & & \\
\hline
 $\tilde M$	& 0.5058626	&  0.4969855		&  &  &  &  &  & & & \\
\hline
 $\tilde u_1$			& 0.1972407 & 0.611379 &  &  &  &  &  & & & \\
\hline
 $\tilde \mu$			& 1.511959577 & 1.1067708306 &  &  &  &  &  & & & \\
\hline
\multicolumn{11}{|c|}{$f_2 = - 0.03$} \\
\hline
 $\tilde E$	& 0.99 		&  				&  &  &  &  &  & & & \\
\hline
 $\tilde W_t$			& 128.514 &  				&  &  &  &  &  & & &  \\
\hline
 $\tilde M$	& 0.4974293	&  		&  &  &  &  &  & & & \\
\hline
 $\tilde u_1$			& 0.6127245 &  &  &  &  &  &  & & & \\
\hline
 $\tilde \mu$			& 1.344502706 &  &  &  &  &  &  & & & \\
\hline
\multicolumn{11}{|c|}{$f_2 = - 0.055$} \\
\hline
 $\tilde E$	& 0.999 & 0.99 & 0.98 &  &  &  &  & &  & \\
\hline
 $\tilde W_t$			& 119.648 & 107.44 & 100.949 &  &  &  & & & &  \\
\hline
 $\tilde M$	& 0.516196205	&  0.50543174		& 0.4936347 &  &  &  &  & & & \\
\hline
 $\tilde u_1$			& 0.243579 & 0.523102 & 0.84453462 &  &  &  &  & & & \\
\hline
 $\tilde \mu$			&1.5327741656  &1.51099935  &1.11405352  &  &  &  &  & & & \\
\hline
\multicolumn{11}{|c|}{$f_2 = - 0.1$} \\
\hline
 $\tilde E$	& 0.99 & 0.96 &  &  &  &  & & &  & \\
\hline
 $\tilde W_t$			& 77.3746 & 85.6755 &  &  &  &  & & &  & \\
\hline
 $\tilde M$	& 0.521154377	&  0.48572		&  &  &  &  &  & & & \\
\hline
 $\tilde u_1$			& 0.494288 & 1.13951853 &  &  &  &  &  & & & \\
\hline
 $\tilde \mu$			& 1.541970066 & 0.3700205133 &  &  &  &  &  & & & \\
\hline
\multicolumn{11}{|c|}{$f_2 = - 0.2$} \\
\hline
 $\tilde E$	& 0.99 & 0.95 & 0.94 & 0.935 &  &  & & &  & \\
\hline
 $\tilde W_t$			& 53.4363 & 72.8364 & 75.9037 & 77.2632 &  &  & & &  & \\
\hline
 $\tilde M$	& 0.543155205	&  0.486725		& 0.4803 & 0.47755 &  &  &  & & & \\
\hline
 $u_1$			& 0.5185539 & 1.22915 & 1.3384 & 1.382189 &  &  &  & & & \\
\hline
 $\tilde \mu$			& 1.58168425006 & 1.28706054 & 0.93349082 & 0.65949247 &  &  &  & & & \\
\hline
\multicolumn{11}{|c|}{$f_2 = - 0.3$} \\
\hline
 $\tilde E$	& 0.99 & 0.95 & 0.92 &  &  &  & & &  & \\
\hline
 $\tilde W_t$			& 43.7621 & 53.2165 & 71.4994 &  &  &  & & &  & \\
\hline
 $\tilde M$	& 0.560506893	&  0.53339		& 0.4761 &  &  &  &  & & & \\
\hline
 $\tilde u_1$			& 0.5426051 & 0.986012 & 1.4828 &  &  &  &  & & & \\
\hline
 $\tilde \mu$			& 1.611834686 & 1.5618306 & 1.07601345 &  &  &  &  & & & \\
\hline
\multicolumn{11}{|c|}{$f_2 = - 0.5$} \\
\hline
 $\tilde E$	& 0.999 & 0.99 & 0.95 & 0.9 & 0.86 &  & & &  & \\
\hline
 $\tilde W_t$			& 45.0617 & 35.794 &41.5654 & 50.5971 &77.1226 &  & & &  & \\
\hline
 $\tilde M$	& 0.5911	&  0.58888149		& 0.5736 & 0.528397 & 0.45553 &  &  & & & \\
\hline
 $\tilde u_1$			& 0.4228 & 0.579942195 & 0.94907 & 1.354263 & 1.8042 &  &  & & & \\
\hline
 $\tilde \mu$			& 1.664592205 & 1.6594926345 & 1.63061992 & 1.5458587 & 0.611046985 &  &  & & & \\
\hline
\multicolumn{11}{|c|}{$f_2 = - 0.75$} \\
\hline
 $\tilde E$	& 0.999 & 0.99 & 0.95 & 0.9 & 0.8 &  & & &  & \\
\hline
 $\tilde W_t$			& 40.1438 & 32.7476 & 36.797 & 41.5315 & 83.3369 &  & & & &  \\
\hline
 $\tilde M$	& 0.6197	&  0.6179		& 0.6074 & 0.5844 & 0.4397 &  &  & & & \\
\hline
 $\tilde u_1$			& 0.46625 & 0.613945 & 0.95271 & 1.2634 & 2.01221 &  &  & & & \\
\hline
 $\tilde \mu$			& 1.71081065 & 1.70675005 & 1.6859686 & 1.6431833 & 0.42607975 &  &  & & & \\
\hline
\multicolumn{11}{|c|}{$f_2 = - 1$} \\
\hline
 $\tilde E$	& 0.999 & 0.995 & 0.99 & 0.95 & 0.9 & 0.8 & & &  & \\
\hline
 $\tilde W_t$			& 38.4915 & 32.8123 &32.3871 & 35.506 & 38.969 & 49.3437 & & &  &
 \\
\hline
 $\tilde M$	& 0.6441	&  	0.6435	& 0.64278 & 0.63455 & 0.61812 & 0.5405 &  & & & \\
\hline
 $\tilde u_1$			& 0.498393 & 0.5752465 & 0.63976 & 0.9628855 & 1.24783 & 1.757792 &  & & & \\
\hline
 $\tilde \mu$			& 1.74969239 & 1.748021225285 & 1.7462147 & 1.729518888 & 1.698292 & 1.549794 &  & & & \\
\hline
\multicolumn{11}{|c|}{$f_2 = - 1.5$} \\
\hline
 $\tilde E$	& 0.999 & 0.99 & 0.95 & 0.9 & 0.8 &  & & &  & \\
\hline
 $\tilde W_t$			& 39.069 & 34.5479 & 36.6868 & 39.0081 & 43.0979 &  & & &  &  \\
\hline
 $\tilde M$	& 0.686	&  0.68492		& 0.6791 & 0.6687 & 0.6319 &  &  & & & \\
\hline
 $\tilde u_1$			& 0.544545 & 0.67777 & 0.98365 & 1.24726 & 1.65477 &  &  & & & \\
\hline
 $\tilde \mu$			& 1.814320967 & 1.81164013 & 1.799302346 & 1.7787229 & 1.7076946 &  &  & & & \\
\hline
\multicolumn{11}{|c|}{$f_2 = - 2$} \\
\hline
 $\tilde E$	& 0.999 	& 0.99 		& 0.95 		& 0.9 		& 0.8 		& 0.65 & & &  &
 \\
\hline
 $\tilde W_t$			& 41.6129 & 38.6684 & 40.0291 & 41.4517 & 44.1617 & 51.0508 & & & &
 \\
\hline
 $\tilde M$	& 0.7217	&  0.7209038		& 0.716345 & 0.70807 & 0.6824 & 0.5933 &  & & & \\
\hline
 $\tilde u_1$			& 0.57797 & 0.705655& 1.001478 & 1.2549 & 1.63822 & 2.100448 &  & & & \\
\hline
 $\tilde \mu$			& 1.868958245 & 1.866712314 & 1.856733174 & 1.84002302 & 1.78885015 & 1.612653 &  & & & \\
\hline
\multicolumn{11}{|c|}{$f_2 = - 5$} \\
\hline
 $\tilde E$	& 0.999 & 0.99 & 0.95 & 0.9 & 0.85 & 0.8 & 0.6	& 0.4 & 0.3 & 0.285
 \\
\hline
 $\tilde W_t$			& 63.4061 & 61.9859 & 62.5747 & 63.2838& 63.8207 & 64.1784 &
 65.0102 & 65.7632 & 69.6015 & 74.9163 \\
\hline
 $\tilde M$	& 0.8746171	&  0.874197172522		& 0.87220248 & 0.86875 & 0.86437 & 0.859 & 0.82487 & 0.754 & 0.6566& 0.6051\\
\hline
 $\tilde u_1$			& 0.6775593 & 0.7899429 & 1.06192505 & 1.29695 & 1.4866 & 1.64963 &  2.150675 &2.518245 &2.712094 & 2.777179\\
\hline
 $\tilde \mu$			& 2.095709819 & 2.0946025448 & 2.0898233502 & 2.0821753622 & 2.0724943067 & 2.06069132 & 1.98759467 & 1.83474942 & 1.61102212 & 1.46293813\\
\hline
\multicolumn{11}{|c|}{$f_2 = - 10$} \\
\hline
 $\tilde E$	& 0.8 & 0.6 & 0.4 & 0.2 & 0.1 &  & & &  &  \\
\hline
 $\tilde W_t$			& 97.2742 & 97.4473 & 97.2636 & 95.5211 & 94.0984 &  & & &  &  \\
\hline
 $\tilde M$	& 1.0367 & 1.02065 & 0.995 & 0.9477 & 0.90575 &  &  & & & \\
\hline
 $\tilde u_1$			& 1.6732 & 2.17354& 2.5443 & 2.841275 & 2.978204 &  &  & & & \\
\hline
 $\tilde \mu$			& 2.323585885 & 2.287561159 & 2.22761172 & 2.12184476 & 2.0248009 &  &  & & & \\
\hline
\multicolumn{11}{|c|}{$f_2 = - 20$} \\
\hline
 $\tilde E$	  & 0.99& 0.8 & 0.6 & 0.4 & 0.2 & 0.1 &  & & &   \\
\hline
 $\tilde W_t$	 &	150.043	& 150.458 & 150.137 & 150.499 & 149.143 & 148.621 &  & & &    \\
\hline
 $\tilde M$ &1.27872	& 1.2756 & 1.26741 & 1.257 & 1.2383 & 1.22613 &  & & &  \\
\hline
 $\tilde u_1$		 & 0.87715	& 1.68655 & 2.19175 & 2.57973 & 2.898958 & 3.04085 &  & & &  \\
\hline
 $\tilde \mu$	 &	2.6818320333	& 2.6727842881 & 2.655017293 & 2.627081697 & 2.58528345 & 2.5563148 &  & & &  \\
\hline
\end{longtable}

\end{document}